\documentclass[aps,nofootinbib,prd,twocolumn,preprintnumbers]{revtex4-1}
\pdfoutput=1 

\usepackage[T1]{fontenc} 
\usepackage{graphicx}
\usepackage{epsfig}
\usepackage{rotating}
\usepackage{color}
\usepackage[normalem]{ulem}
\usepackage{multirow}
\usepackage{tabularx}
\usepackage{graphicx}
\usepackage{dcolumn}
\usepackage{hyperref}
\usepackage{float}
\usepackage{amsmath}
\usepackage{amssymb}
\usepackage{float}
\usepackage{booktabs}

\usepackage{comment}

\excludecomment{ignore}

\usepackage{ulem}
\usepackage{color}

\begin{document} 
\title{ Novel neutrino-floor and dark matter searches with deformed shell model 
calculations}
\author{D.K. Papoulias~$^{1}$}\email{dimpap@cc.uoi.gr}
\author{R. Sahu~$^2$}\email{rankasahu@gmail.com}
\author{T.S. Kosmas~$^3$}\email{hkosmas@uoi.gr}
\author{V.K.B. Kota~$^4$}\email{vkbkota@prl.res.in}
\author{B. Nayak~$^{2}$}\email{bishnukar@nist.edu}

\affiliation{$^{1}$~Institute of Nuclear and Particle Physics, NCSR `Demokritos', Agia Paraskevi, 15310, Greece}
\affiliation{$^2$~National Institute of Science and Technology, Palur Hills, Berhampur-761008, Odisha, India} 
\affiliation{$^3$~Theoretical Physics Section, University of Ioannina,
  GR-45110 Ioannina, Greece} 
\affiliation{$^4$~Physical Research Laboratory, Ahmedabad 380 009, India} 

\begin{abstract}  
Event detection rates for WIMP-nucleus interactions are calculated for $^{71}$Ga, $^{73}$Ge, $^{75}$As and $^{127}$I (direct dark matter detectors). The nuclear structure form factors, that are rather independent of the underlying beyond the Standard Model particle physics scenario assumed, are evaluated within the context of the deformed nuclear shell model (DSM) based on Hartree-Fock nuclear states. Along with the previously published DSM results for $^{73}$Ge, the neutrino-floor due to coherent elastic neutrino-nucleus scattering (CE$\nu$NS), an important source of background to dark matter searches, is extensively calculated. The impact of new contributions to CE$\nu$NS due to neutrino magnetic moments and $Z^\prime$ mediators at direct dark matter detection experiments is also examined and discussed. The results show that the neutrino-floor constitutes a crucial source of background events for multi-ton scale detectors with sub-keV capabilities. 
\end{abstract}

\maketitle

\section{Introduction}
\label{sect:intro}

In the last few decades, the measurements of the cosmic microwave background (CMB) radiation offered a remarkably powerful way of modelling the origin of cosmic-ray anisotropies and constraining the geometry, the evolution and the matter content of our universe. Such observations have in general indicated the consistency of the standard cosmological model~\cite{Jungman:1995df} and the fact that our universe hardly contains $\sim 5$\% luminous matter, whereas the remainder consists of non-luminous dark matter ($\sim 23$\%) and dark energy ($\sim 72$\%)~\cite{Hinshaw:2012aka}. After the discovery of the CMB fluctuations by the Cosmic Background Explorer (COBE) satellite~\cite{Smoot:1992td}, the extremely high precision of the WMAP satellite and especially of the Planck third-generation space mission, have helped us to produce maps for the CMB anisotropies and other cosmological parameters (see Ref.~\cite{Ade:2015xua} for details). We also mention that, high-resolution ground-based CMB data, like those of the Atacama Cosmology Telescope (ACT)~\cite{Das:2013zf} and the South Pole Telescope (SPT)~\cite{George:2014oba} have recovered the underlying CMB spectra observed by the space missions. 

Focusing on the topic we address in this work, it is worth noting that the CMB data, the Supernova Cosmology project~\cite{Gawiser:1998zh}, etc., suggest that most of the dark matter of the universe is cold. Furthermore, the baryonic cold dark matter (CDM) component can be considered to consist of either massive compact halo objects (MACHOs) like neutron stars, white dwarfs, jupiters, etc., or Weakly Interacting Massive Particles (WIMPs) that constantly bombard Earth's atmosphere. Several results of  experimental searches suggest that the MACHO fraction should not exceed a portion of about 20\%~\cite{Jungman:1995df}. On the theoretical side, within the framework of new physics beyond the Standard Model (SM), supersymmetric (SUSY) theories provide promising non-baryonic candidates for dark matter~\cite{Kosmas:1997jm} (for a review see Ref.~\cite{Bertone:2004pz}). In the simple picture, the dark matter in the galactic halo is assumed to be Weakly Interacting Massive Particles (WIMPs). The most appealing WIMP candidate for non-baryonic CDM is the lightest supersymmetric particle (LSP) which is expected to be a neutral Majorana fermion traveling with non-relativistic velocities~\cite{Kortelainen:2006rd}. 

In recent years, there have been considerable theoretical and experimental efforts towards WIMP detection through several nuclear probes~\cite{Feng:2010gw,Fitzpatrick:2012ix,Vergados:2015tua}. Popular target nuclei include  among others the $^{71}$Ga, $^{73}$Ge, $^{75}$As, $^{127}$I, $^{134}$Xe and $^{208}$Pb isotopes~\cite{Divari:2000dc,Holmlund:2004rv}. Towards the first ever dark matter detection, a great number of experimental efforts take place aiming at measuring the energy deposited after the galactic halo WIMPs scatter off the nuclear isotopes of the detection material. Because of the low count rates, due to the fact that the WIMP-nucleus interaction is remarkably weak, the choice of the detector plays very important role and for this reason spin-dependent interactions require the use of targets with non-zero spin. The Cryogenic Dark Matter Search (CDMS) experimental facility~\cite{Akerib:2005zy} has been designed to directly detect the dark matter by employing a $^{73}$Ge as the target nucleus, setting the most sensitive limits on the interaction of WIMP with terrestrial materials to date. The development of upgrades is under way and will be located at SNOLAB. Another prominent dark matter experiment is the EDELWEISS facility in France~\cite{Broniatowski:2009qu} which uses high purity germanium cryogenic bolometers at millikelvin temperatures.  There are also other experimental attempts using detectors like  $^{127}$I, $^{129,131}$Xe, $^{133}$Cs, etc. (see Refs.~\cite{Freese:2012xd, Akerib:2016vxi,Fu:2016ega}). 

Inevitably, direct detection experiments are exposed to various neutrino emissions, such as those originating from astrophysical sources (e.g. Solar~\cite{Robertson:2012ib}, Atmospheric~\cite{Battistoni:2005pd,Ng:2017aur} and diffuse Supernova~\cite{Beacom:2010kk} neutrinos), Earth neutrinos (Geoneutrinos~\cite{Monroe:2007xp}), and in other cases even from artificial terrestrial neutrinos (e.g. neutrinos from nearby reactors~\cite{Gelmini:2018ogy}). The subsequent neutrino interactions with the material of dark matter detectors, namely the neutrino-floor, may perfectly mimic possible WIMP signals~\cite{Billard:2013qya}. Thus, the impact of the neutrino-floor on the relevant experiments looking for CDM as well as on the detector responses to neutrino interactions need be comprehensively investigated. Since Geoneutrino fluxes are relatively low, astrophysical neutrinos are recognised as the most significant background source that remains practically irreducible~\cite{OHare:2016pjy}. The recent advances of direct detection dark matter experiments,  mainly due to the development of low threshold technology and high detection efficiency, are expected to reach the sensitivity frontiers in which astrophysical neutrino-induced backgrounds are expected to limit the observation potential of the 
WIMP signal~\cite{Battat:2016pap}. 

In this work, we explore the impact of the most important neutrino background source on the relevant direct dark matter detection experiments by concentrating on the dominant neutrino-matter interaction channel, e.g. the coherent elastic neutrino-nucleus scattering (CE$\nu$NS)~\cite{Freedman:1973yd,Bednyakov:2018mjd}. It is worthwhile to mention that, events of this process were recently measured for the first time by the COHERENT experiment at the Spallation Neutrino Source~\cite{Akimov:2017ade}, completing the SM picture of electroweak interactions at low energies. Such a profound discovery  motivated our present work and we will make an effort to shed light on the nuclear physics aspects. Neutrino non-standard interactions (NSIs)~\cite{Papoulias:2013gha} may constitute an important source of neutrino background and have been investigated recently in Refs.~\cite{Coloma:2017ncl,AristizabalSierra:2017joc}. Thus, apart from addressing the SM contributions to CE$\nu$NS~\cite{Papoulias:2015vxa}, we also explore the impact of new physics contributions that arise in the context of electromagnetic (EM) neutrino properties~\cite{Kosmas:2015sqa,Kosmas:2015vsa} as well as of those emerging in the framework of $U(1)^\prime$ gauge interactions~\cite{Bertuzzo:2017tuf} due to the presence of new light $Z^\prime$ mediators~\cite{Dent:2016wor,Kosmas:2017tsq}. The aforementioned interaction channels may lead to a novel neutrino-floor as demonstrated by Ref.~\cite{Dutta:2015vwa}. The latter could be detectable in view of the constantly increasing sensitivity of the upcoming direct detection experiments with multi-ton mass scale and sub-keV capabilities~\cite{Strigari:2016ztv}. 

Direct detection dark matter experiments are currently entering a precision era, and nuclear structure effects are expected to become rather important and should be incorporated in astroparticle physics applications~\cite{Gardner:2013ama}. For this reason, our nuclear model is at first tested in its capabilities to adequately describe the nuclear properties before being applied to problems like dark matter detection. This work considers the deformed shell model (DSM), on the basis of Hartree-Fock (HF) deformed intrinsic states with angular momentum projection and band mixing~\cite{Kosmas:2003xr}, all with a realistic effective interaction and a set of single particle states and single particle energies, which is established to be rather successful in describing the properties of nuclei in the mass range $A$=60--90 (see Ref.~\cite{ks-book} for details regarding DSM and its applications). In particular, the DSM is employed for calculating the required nuclear structure factors entering the dark matter and neutrino-floor expected event rates by focusing on four interesting nuclei regarding dark matter investigations such as $^{71}$Ga, $^{73}$Ge, $^{75}$As and $^{127}$I. Let us add that details of  nuclear structure and dark matter event rates for $^{73}$Ge obtained using DSM have been reported recently~\cite{Sahu:2017czz}.
 
The paper has been organised as follows: Section~\ref{sect:DM} gives the main ingredients of WIMP-nucleus scattering, while  Sect.~\ref{sect:neutrinos} provides the formulation for neutrino-nucleus scattering (neutrino-floor) within and beyond the SM. Then in Sect.~\ref{sect:DSM} we describe briefly the methodology of the DSM, and the main results of the present work are presented and discussed in Sect.~\ref{sect:results}. Finally, the concluding remarks are drawn in Sect.~\ref{sect:conclusions}.

\section{Searching WIMP Dark Matter}
\label{sect:DM}

The Earth is exposed to a huge number of WIMPs originating from the galactic halo. Their direct detection through nuclear recoil measurements after scattering off the target nuclei at the relevant dark matter experiments is of fundamental interest in modern physics and is expected to have a direct impact to astroparticle physics and cosmology. In this section we discuss the mathematical formulation of WIMP-nucleus scattering. The formalism introduces an appropriate separation of the SUSY and nuclear parts entering the event rates  of WIMP-nucleus interactions in our effort to emphasise the important  role played by the nuclear physics aspects. In particular we perform reliable nuclear structure calculations within the context of DSM based on Hartree-Fock states.

\subsection{WIMP-nucleus scattering}
For direct detection dark matter experiments, the differential event rate of a WIMP with mass $m_\chi$ scattering off a nucleus $(A,Z)$ with respect to the momentum transfer $q$, can be cast in the form~\cite{Jungman:1995df}
\begin{equation}
\frac{dR(u,\upsilon)}{dq^2} = N_t \phi \frac{d \sigma}{d q^2} f(\upsilon) \, d^3 \upsilon \, ,
\label{eq:DM-rate}
\end{equation}
where, $N_t = 1/( A m_p)$ denotes the number of target nuclei per unit mass, $A$ stands for the mass number of the target nucleus and $m_p$ is the proton  mass. In the above expression the WIMP flux is $\phi=\rho_0 \upsilon/m_\chi$, with $\rho_0$ being the local WIMP density. The distribution of WIMP velocity relative to the detector (or Earth) and also the motion of the Sun and Earth, $f(\upsilon)$, is taken into account and assumed to resemble a Maxwell-Boltzmann distribution to ensure consistency with the LSP velocity distribution. Note that, by neglecting the rotation of Earth  in its own axis, then $\upsilon= \vert \mathbf{v}\vert$ accounts for  the relative velocity of WIMP with respect to the detector. For later convenience a dimensionless variable $u=q^2b^2 /2$ is introduced with $b$ denoting the oscillator length parameter, and the corresponding WIMP-nucleus differential cross section in the laboratory frame reads~\cite{Pirinen:2016pxr,Toivanen:2009zza,Kortelainen:2006rd,Holmlund:2004rv,
Sahu:2017czz}
\begin{equation}
\frac{d \sigma (u, \upsilon)}{d u} = \frac{1}{2} \sigma_0 \left(\frac{1}{m_p b} \right)^2 \frac{c^2}{\upsilon^2} \frac{d \sigma_A (u)}{du} \, ,
\end{equation}
with
\begin{equation}
\begin{aligned}
\frac{d \sigma_A}{du} =&  \left[f_A^0 \Omega_0(0) \right]^2 F_{00}(u) \\  & +2 f_A^0 f_A^1 \Omega_0 (0) \Omega_1(0) F_{01}(u) \\  & + \left[f_A^1 \Omega_1(0) \right]^2 F_{11}(u) + \mathcal{M}^2(u) \, .
\end{aligned}
\label{eq:DM-crossec}
\end{equation}
The first three terms account for the spin  contribution due to the axial current, while the fourth term accounts for the coherent contribution arising from the scalar interaction.  The coherent contribution is expressed in terms of the nuclear form factors given as
\begin{equation}
\begin{aligned}
\mathcal{M}^2(u) = & \bigl( f_S^0 \left[Z F_Z(u) + N F_N(u) \right]\\ 
& + f_S^1 \left[ Z F_Z(u) - N F_N(u)\right] \bigr)^2 \, .
\end{aligned}
\end{equation}
The coherent part in the approximation of nearly equal proton and neutron nuclear form factors $F_Z(u) \approx F_N(u)$, is given as
\begin{equation}
\mathcal{M}^2(u) = A^2 \left(f_S^0 - f_S^1 \frac{A-2Z}{A} \right)^2 \vert F(u) \vert^2 \, .
\end{equation}
The respective values of the nucleonic-current parameters $f_V^0$, $f_V^1$ for 
the isoscalar and isovector parts of the vector current (not shown here), 
$f_A^0$, $f_A^1$ for the isoscalar and isovector parts of the axial-vector 
current, and $f_S^0$, $f_S^1$ for the isoscalar and isovector parts of the 
scalar current, depend on the specific SUSY model 
employed~\cite{Vergados:1998jt}. 
The spin structure functions $F_{\rho \rho^\prime}(u)$ with $\rho , \, \rho^\prime = 0,1$ for the isoscalar and isovector contributions respectively, take the form
\begin{equation}
F_{\rho \rho^\prime}(u) = \sum_{\lambda, \kappa}  \frac{\Omega_{\rho}^{(\lambda,\kappa)}(u) \Omega_{\rho^\prime}^{(\lambda,\kappa)}(u)}{\Omega_\rho(0) \Omega_{\rho^\prime}(0)}\, ,
\label{eq:spin-struct1}
\end{equation}
with
\begin{equation}
\begin{aligned}
& \Omega_{\rho}^{(\lambda,\kappa)}(u) = \sqrt{\frac{4 \pi}{2 J_i +1}} \\ &\times \langle J_f \vert \vert \sum_{j=1}^{A} \left[Y_\lambda (\Omega_j) \otimes \sigma(j) \right]_\kappa j_\lambda(\sqrt{u} \,  r_j) \omega_\rho(j) \vert \vert J_i \rangle \, .
\end{aligned}
\label{eq:spin-struct2}
\end{equation}
Here, $\omega_0(j) = 1$ and $\omega_1(j) = \tau(j)$ with  $\tau = +1$ for protons and $\tau = -1$ for neutrons, while $\Omega_j$ represents the solid angle for the position vector of the $j$-th nucleon and $j_\lambda$ stands the well-known spherical Bessel function. The quantities $\Omega_\rho (0) = 
\Omega_\rho^{(0,1)}(0)$ are the static spin matrix elements 
(see e.g. Ref.~\cite{Kosmas:1997jm}). In this context, the WIMP-nucleus event rate per unit mass of the detector is conveniently written as
\begin{equation}
\begin{aligned}
\langle R \rangle =& (f_A^0)^2 D_1 + 2 f_A^0 f_A^1 D_2 + (f_A^1)^2 D_3\\ & + A^2 \left(f_S^0 - f_S^1  \frac{A-2Z}{A}\right)^2 \vert F(u)\vert^2 D_4\, .
\end{aligned}
\end{equation}
The functions $D_i$ enter the definition of the WIMP-nucleus event rate through the three-dimensional integrals, given by
\begin{equation}
D_i = \int_{-1}^1 d \xi \int_{\psi_{\mathrm{min}}}^{\psi_{\mathrm{max}}} d \psi \int_{u_{\mathrm{min}}}^{u_{\mathrm{max}}} G(\psi, \xi) X_i \, du\, ,
\label{eq:DM-int2}
\end{equation}
with
\begin{equation}
\begin{aligned}
X_1 =& \left[\Omega_0(0) \right]^2 F_{00}(u)\, , \\
X_2 =& \Omega_0 (0) \Omega_1(0) F_{01}(u)\,  , \\
X_3 =& \left[\Omega_1(0) \right]^2 F_{11}(u)\,  , \\
X_4 =& \vert F(u) \vert^2\, .
\end{aligned}
\end{equation}
In the latter expression, $D_1$, $D_2$, $D_3$ account for the spin dependent parts of 
Eq.(\ref{eq:DM-crossec}) while  $D_4$ is associated to the coherent 
contribution. 

In this work, the nuclear wave functions $\langle J_f \vert$ and $ \vert J_i \rangle$ entering
Eq.(\ref{eq:spin-struct2}) are calculated within the nuclear DSM of Refs.~\cite{Kosmas:2003xr}
and~\cite{ks-book}.
For a comprehensive discussion on the explicit form of the function $G(\psi)$, the integration limits of 
Eq.(\ref{eq:DM-int2}) and the various parameters entering into these, the reader is referred to Ref.~\cite{Sahu:2017czz}. 

\section{Neutrino-Nucleus Scattering}
\label{sect:neutrinos}

The neutrino-floor stands out as an important source of irreducible background to WIMP searches at a direct detection experiment. In this work we explore the neutrino-floor due to neutrino-nucleus scattering since the corresponding floor coming from neutrino-electron scattering is relatively low~\cite{Wyenberg:2018eyv}. Motivated by the novel neutrino interaction searches using reactor neutrinos of Ref.~\cite{Dutta:2015vwa}, here we consider various astrophysical neutrino sources in our calculations that involve the conventional and beyond the SM interactions channels (see below).

\subsection{Differential event rate at dark matter detectors}
 For a given interaction channel $x={\mathrm{SM, EM}, Z^\prime} $, the differential event rate $d R_\nu/dT_N$ of  CE$\nu$NS processes at a dark matter detector is obtained through the convolution of the normalised neutrino energy distribution $\lambda_{\nu}(E_\nu)$ of the background neutrino source in question (i.e. Solar, Atmospheric and Diffuse 
Supernova Neutrinos, see below) with the CE$\nu$NS cross section, as follows~\cite{Papoulias:2015iga}
\begin{equation}
\left(\frac{d R_\nu}{dT_N}\right)_{\mathrm{x}} =  \mathcal{K}  \int_{E_{\nu}^{\mathrm{min}}}^{E_{\nu}^{\mathrm{max}}} \lambda_{\nu}(E_\nu)  \frac{d \sigma_{\mathrm{x}}}{dT_N} (E_\nu, T_N) \, dE_\nu  \, ,
\label{eq:event-rate}
\end{equation}
where $E_{\nu}^{\mathrm{max}}$ the maximum neutrino energy of the source in question (for the case of Solar neutrinos see e.g. Table~\ref{table:fluxes}) and $E_{\nu}^{\mathrm{min}}=\sqrt{M T_N/2}$ is the minimum neutrino energy that is required to yield a nuclear recoil with energy $T_N$. In the latter expression $\mathcal{K} =  t_{\mathrm{run}} N_{\mathrm{targ}} \Phi_\nu$ with $t_{\mathrm{run}}$ being the exposure time, $N_{\mathrm{targ}}$ is the number of target nuclei and $\Phi_\nu$ is the assumed neutrino flux.

\subsubsection{Standard Model interactions}
Assuming SM interactions only, at low and intermediate neutrino energies $E_\nu \ll M_W$, the weak neutral-current CE$\nu$NS process is adequately described by the four-fermion effective interaction Lagrangian~\cite{Papoulias:2013gha,Papoulias:2015vxa} 
\begin{equation}
\mathcal{L}_{\mathrm{SM}} = - 2 \sqrt{2} G_{F} \sum_{ \begin{subarray}{c} f= \, u,d \\ \alpha= e, \mu,\tau 
\end{subarray}} g_{\alpha \alpha}^{f,P}   
\left[ \bar{\nu}_{\alpha} \gamma_{\rho} L \nu_{\alpha} \right] \left[ \bar{f} \gamma^{\rho} P f \right] \, ,
\label{SM_Lagr}
\end{equation}
where $P=\{L,R\}$ denote the chiral projectors, $\alpha=\{e,\mu,\tau \}$ represents the neutrino flavour and $f=\{u,d\}$ is a first generation quark. By including the radiative corrections of Ref.~\cite{Beringer:1900zz}, the $P$-handed couplings of the $f$ quarks to the $Z$-boson, are expressed as 
\begin{equation}
\begin{aligned}
g_{\alpha \alpha}^{u,L} =& \rho_{\nu N}^{NC} \left( \frac{1}{2}-\frac{2}{3} \hat{\kappa}_{\nu N} \hat{s}^2_Z \right) + \lambda^{u,L} \, ,\\
g_{\alpha \alpha}^{d,L} =& \rho_{\nu N}^{NC} \left( -\frac{1}{2}+\frac{1}{3} \hat{\kappa}_{\nu N} \hat{s}^2_Z \right) + \lambda^{d,L} \, ,\\
g_{\alpha \alpha}^{u,R} =& \rho_{\nu N}^{NC} \left(-\frac{2}{3} \hat{\kappa}_{\nu N} \hat{s}^2_Z \right) + \lambda^{u,R} \, ,\\
g_{\alpha \alpha}^{d,R} =& \rho_{\nu N}^{NC} \left(\frac{1}{3} \hat{\kappa}_{\nu N} \hat{s}^2_Z \right) + \lambda^{d,R} \, ,
\end{aligned}
\end{equation}
with $\hat{s}^2_Z = \sin^2 \theta_W= 0.2312$, $\rho_{\nu N}^{NC} = 1.0086$, $\hat{\kappa}_{\nu N} = 0.9978$, $\lambda^{u,L} = -0.0031$, $\lambda^{d,L} = -0.0025$ and $\lambda^{d,R} =2\lambda^{u,R} = 7.5 \times 10^{-5}$.

In this work we restrict our study only to low momentum transfer in order to satisfy the coherent condition $\vert \mathbf{q} \vert \leq 1/R_A$, where $R_A$ is the nuclear size and $\vert \mathbf{q} \vert$ is the magnitude of the three-momentum transfer~\cite{Bednyakov:2018mjd}. Focusing on the dominant CE$\nu$NS channel, the relevant SM differential cross section with respect to the nuclear recoil energy $T_N$, takes the form~\cite{Kosmas:2017tsq}
\begin{equation}
\begin{aligned}
\frac{d \sigma_{\mathrm{SM}}}{dT_N}(E_\nu, T_N) = &\frac{G_F^2 M}{\pi}  \Biggl[ (\mathcal{Q}_W^V)^2  \left(1 - \frac{M T_N}{2 E_\nu^2} \right)\\  &+ (\mathcal{Q}_W^A)^2  \left(1 + \frac{M T_N}{2 E_\nu^2} \right) \Biggr] \, ,
\end{aligned}
\label{eq:diff-crossec}
\end{equation}
with $E_\nu$ denoting the neutrino energy and $M$ the mass of the target nucleus. The relevant vector ($\mathcal{Q}_W^V$) and axial-vector ($\mathcal{Q}_W^A$) weak charges entering the CE$\nu$NS cross section, are given by the relations~\cite{Barranco:2005yy}
\begin{equation}
\begin{aligned}
\mathcal{Q}_W^V(Q^2) =& \left[ g^V_p Z F_Z^V(Q^2)+ g^V_n N F_N^V(Q^2)\right]  \, , \\
\mathcal{Q}_W^A (Q^2) =& \left[ g^A_p (Z_+ - Z_-) + g^A_n (N_+ - N_-) \right] F_A(Q^2)\, .
\end{aligned}
\label{eq:weak-charge}
\end{equation}
Here, $Z_{\pm}$ ($N_{\pm}$) stands for the number of protons (neutrons) with spin up ($+$) and spin down ($-$), respectively, while $g^A_p$ ($g^A_n$) represent for the axial-vector couplings of protons (neutrons) to the $Z^0$ boson. At the nuclear level, the relevant vector (axial-vector) couplings of protons $g^V_p$ ($g^A_p$) and neutrons  $g^V_n$ ($g^A_n$) take the form 
\begin{equation}
\begin{aligned}
g^V_p =  & 2(g_{\alpha \alpha}^{u,L} + g_{\alpha \alpha}^{u,R}) + (g_{\alpha \alpha}^{d,L} + g_{\alpha \alpha}^{d,R}) \, , \\
g^V_n = &(g_{\alpha \alpha}^{u,L} +
g_{\alpha \alpha}^{u,R}) +2(g_{\alpha \alpha}^{d,L} + g_{\alpha
  \alpha}^{d,R}) \, , \\
g^A_p =  & 2(g_{\alpha \alpha}^{u,L} - g_{\alpha \alpha}^{u,R}) + (g_{\alpha \alpha}^{d,L} - g_{\alpha \alpha}^{d,R}) \, , \\
g^A_n = &(g_{\alpha \alpha}^{u,L} -
g_{\alpha \alpha}^{u,R}) +2(g_{\alpha \alpha}^{d,L} - g_{\alpha
  \alpha}^{d,R})  \, .
\end{aligned}
\end{equation}
The axial vector nucleon form factor takes into account the spin structure of the nucleon and is defined as~\cite{Alberico:2001sd}
\begin{equation}
F_A(Q^2)= g_A \left(1 + \frac{Q^2}{M_A^2}  \right)^{-2} \, ,
\end{equation}
where $g_A=1.267$ is the free axial-vector coupling constant and the axial mass is taken to be $M_A=1$~GeV, while strange quark effects have been neglected.

We note that for spin-zero nuclei the axial-vector contribution vanishes, while for the odd-$A$ nuclei considered in the present study $\mathcal{Q}_W^A$ it is negligible and of the order of $\mathcal{Q}_W^A/\mathcal{Q}_W^V \sim 1/A$.  The weak charges in Eq.(\ref{eq:weak-charge}) encode crucial information regarding the finite nuclear size through the proton $F_Z^V(Q^2)$ and neutron $F_N^V(Q^2)$ nuclear form factors, which in our work are obtained within the context of the DSM  (see below), as functions of the momentum transfer $-q^\mu q_\mu=Q^2 = 2 M T_N$. Contrary to similar studies assuming the conventional Helm-type form factors, the present work also takes into account the nuclear effects due to the non-spherical symmetric nuclei employed in dark matter searches.

\subsubsection{Electromagnetic neutrino contributions}
Turning our attention to new physics phenomena we now address potential contributions to CE$\nu$NS in the framework of non-trivial neutrino EM interactions, that may lead to a new neutrino-floor at low detector thresholds. In this framework, the presence of an effective neutrino magnetic moment $\mu_\nu$, leads an EM contribution of the differential cross section, that has been written as~\cite{Kosmas:2017tsq}
\begin{equation}
\left(\frac{d \sigma}{dT_N} \right)_{\mathrm{SM+EM}} = \mathcal{G_{\mathrm{EM}}}(E_\nu, T_N) \frac{d \sigma_{\mathrm{SM}}}{d T_N} \, . 
\label{eq:EM-crossec}
\end{equation}
Neglecting axial effects, the EM contribution to CE$\nu$NS at a direct detection dark matter is encoded in the factor
\begin{equation}
\mathcal{G}_\mathrm{EM} = 1 + \frac{1}{G_F^2 M}\left(\frac{\mathcal{Q}_{\mathrm{EM}}}{\mathcal{Q}_W^V} \right)^2 \frac{\frac{1- T_N/E_\nu}{T_N}}{1 - \frac{M T_N}{2 E_\nu^2}} \, ,
\label{EM:factor}
\end{equation}
where, the relevant EM charge $\mathcal{Q}_{\mathrm{EM}}$ is written in terms of the electron mass $m_e$, the fine-structure constant $a_{\mathrm{EM}}$ and the effective neutrino magnetic moment as~\cite{Scholberg:2005qs}
\begin{equation}
\mathcal{Q}_{\mathrm{EM}} = \frac{\pi a_{\mathrm{EM}} \mu_{\nu}}{m_e} Z \, .
\label{eq:EM-charge}
\end{equation}
In contrast to the $\sim N^2$ dependence of the SM case, Eqs. (\ref{EM:factor}) and (\ref{eq:EM-charge}) imply the existence of a $Z^2$ coherence along with a characteristic $\sim 1/T_N$ enhancement of the total cross section. This implies a potential distortion of the expected recoil spectrum at very low recoil energies that may be detectable at future direct dark matter detection with sub-keV operation thresholds. 

For the sake of completeness we stress that the effective neutrino magnetic moment $\mu_\nu$, is expressed through neutrino
amplitudes of positive and negative helicity states, e.g. the 3-vectors $a_+$ and $a_-$ and the neutrino transition magnetic moment matrix, $\lambda$, in flavour basis, as~\cite{Grimus:2002vb, Kosmas:2015sqa}
\begin{equation}
\mu_\nu^2= a^\dag_+ \lambda \lambda^\dag a_+ + a^\dag_- \lambda \lambda^\dag a_-\, .
\end{equation}
Then, the effective neutrino magnetic moment is written in mass basis through a proper rotation, for a detailed description of this formalism see Ref.~\cite{Canas:2015yoa}.

\subsubsection{Novel mediator contribution}
We now explore novel mediator fields that could be accommodated in the context of simplified $U(1)^\prime$ scenarios~\cite{Dent:2016wcr,Shoemaker:2017lzs} predicting the existence of a new $Z^\prime$ vector mediator with mass $M_{Z^\prime}$~\cite{Lindner:2016wff}. Such beyond the SM interactions  may constitute a new neutrino-floor at direct detection dark matter experiments~\cite{Bertuzzo:2017tuf}. 

The presence of a $Z^\prime$ mediator gives rise to sub-leading contributions to the SM CE$\nu$NS rate, described by the Lagrangian~\cite{Cerdeno:2016sfi}
\begin{equation}
\mathcal{L}_{\mathrm{vec}} =  Z^{\prime}_\mu \left(g_{Z^\prime}^{qV} \bar{q} \gamma^\mu q + g_{Z^\prime}^{\nu V} \bar{\nu}_L \gamma^\mu \nu_L\right) + \frac{1}{2} M_{Z^\prime}^2 Z^{\prime}_\mu Z^{\prime \mu} \, , 
\label{lagr:z-prime}
\end{equation}
where only left-handed neutrinos are assumed (right-handed neutrinos in the theory would lead to vector-axial-vector cancellations). The resulting cross section reads~\cite{Kosmas:2017tsq}
\begin{equation}
\left( \frac{d \sigma}{dT_N}\right)_{\mathrm{SM} + Z^\prime} = \mathcal{G}_{Z^\prime}^2 (Q) \frac{d \sigma_{\mathrm{SM}}}{d T_N} \, ,
\end{equation}
with the factor $\mathcal{G}_{Z^\prime}$ being written in terms of the neutrino-vector coupling $g_{Z^\prime}^{\nu V}$, as
\begin{equation}
\mathcal{G}_{Z^\prime}(Q) = 
1 - \frac{1}{2 \sqrt{2}G_F}\frac{\mathcal{Q}_{Z^\prime}}{\mathcal{Q}_W^V} \frac{g_{Z^\prime}^{\nu V}}{Q^2 + M_{Z^\prime}^2} \, .
\label{eq:G_z-prime}
\end{equation}
The relevant charge in this case is expressed through the vector  quark couplings $^{qV}_{Z^\prime}$ to the $Z^\prime$ boson, as~\cite{Bertuzzo:2017tuf}
\begin{equation}
\mathcal{Q}_{Z^\prime} = \left(2 g^{uV}_{Z^\prime} + g^{dV}_{Z^\prime} \right)Z +  \left(g^{uV}_{Z^\prime} + 2g^{dV}_{Z^\prime} \right) N \, .
\end{equation}
Let us mention that emerging degeneracies can be either reduced through multi-detector measurements~\cite{Shoemaker:2017lzs} and broken in the framework of NSIs~\cite{Liao:2017uzy}. For completeness we note that, despite being not present for the low energies considered here, these couplings could be changed by currently unknown in-medium effects (see e.g. Ref.~\cite{Ng:2017aur} and references therein).

\begin{figure}[t]
\centering
\includegraphics[width= \linewidth]{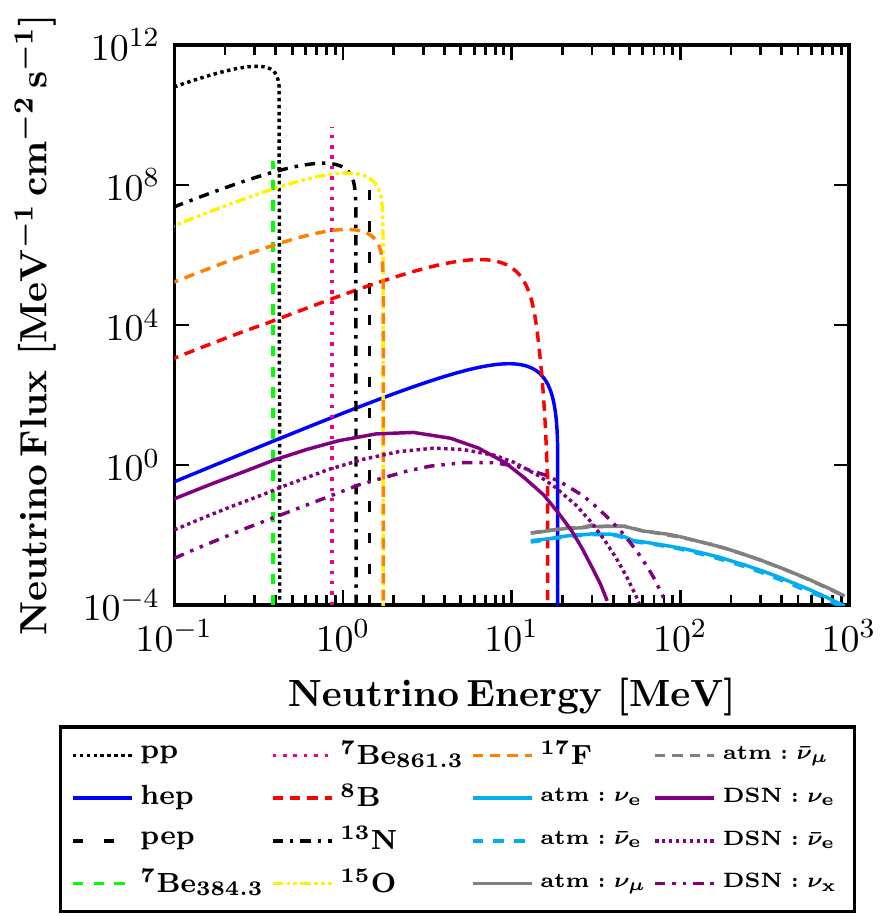}
\caption{Unoscillate neutrino flux considered in the present study, including the Solar, Atmospheric and DSNB spectra.}
\label{fig:solar-spectrum}
\end{figure}
\begin{table}[t]
\begin{tabularx}{\linewidth}{XXX}
\toprule
type & $E_{\nu_{\mathrm{max}}}$ [MeV] & flux [$\mathrm{cm^{-2} s^{-1}}$]\\
\midrule 
$pp$ & 0.423 & $(5.98 \pm 0.006) \times 10^{10}$\\ 
$pep$ & 1.440 & $(1.44 \pm 0.012) \times 10^{8}$\\ 
$hep$ & 18.784 & $(8.04 \pm 1.30) \times 10^{3}$\\
$\mathrm{^7Be_{low}}$ & 0.3843 & $(4.84 \pm 0.48) \times 10^{8}$\\
$\mathrm{^7Be_{high}}$ & 0.8613 & $(4.35 \pm 0.35) \times 10^{9}$\\
$\mathrm{^8B}$ & 16.360 & $(5.58 \pm 0.14) \times 10^{6}$\\
$\mathrm{^{13}N}$ & 1.199 & $(2.97 \pm 0.14) \times 10^{8}$\\
$\mathrm{^{15}O}$ & 1.732 & $(2.23 \pm 0.15) \times 10^{8}$\\
$\mathrm{^{17}F}$ & 1.740 & $(5.52 \pm 0.17) \times 10^{6}$\\
\bottomrule
\end{tabularx}
\caption{Solar neutrino fluxes and uncertainties in the framework of the employed high metallicity SSM (for details, see the text).}
\label{table:fluxes}
\end{table}

\subsection{Neutrino sources}

\subsubsection{Solar Neutrinos}
In terrestrial searches for dark matter candidates at low energies, the Solar neutrinos emanating from the interior of the Sun generated through various fusion reactions produce a dominant background for direct CDM detection experiments.
Assuming WIMP masses less than 10~GeV, an estimated total Solar neutrino flux of about $6.5 \times 10^{11}~\mathrm{cm^{-2} s^{-1}}$~\cite{Antonelli:2012qu} hitting the Earth is expected to appreciably limit the sensitivity of such experiments~\cite{Billard:2013qya}. On the other hand, the theoretical uncertainties of Solar neutrinos are presently quite large and depend strongly on the assumed Solar neutrino model. To maintain consistency with existing Solar data, in this work we consider the high metallicity Standard Solar Model (SSM)~\cite{Robertson:2012ib}. We note however that, the dominant Solar neutrino component coming from the primary proton-proton channel ($pp$ neutrinos) that accounts for about 86\% the Solar neutrinos flux, has been recently measured by the Borexino experiment with an uncertainty of 1\%~\cite{Bellini:2014uqa}. Through CE$\nu$NS, the direct detection dark matter experiments are mainly sensitive to two sources of Solar neutrinos, namely the $^8$B and the $hep$ neutrinos which cover the highest energy range of the Solar neutrino spectrum. Since, $^8$B neutrinos are generated from the decay $\mathrm{^8B} \rightarrow \mathrm{^7Be^*}  + e^+ + \nu_e$ while $hep$ neutrinos from $\mathrm{^3 He} + p \rightarrow \mathrm{^4 He} + e^+ + \nu_e$, both sources occur in the aftermath of the $pp$ chain. Following previous similar studies~\cite{OHare:2016pjy}, in this work, we explore the neutrino-floor extending our analysis to the lowest neutrino energies, by considering the $pep$ neutrino line which belongs to the $pp$ chain and the $e^-$-capture reaction on $^7$Be that leads to two monochromatic beams at 384.3 and 861.3~keV as well as the well known CNO cycle. The latter neutrinos appear as three continuous spectra ($^{13}$N, $^{15}$O, $^{17}$F) with end point energies close to the $pep$ neutrinos.

\subsubsection{Atmospheric Neutrinos}

Atmospheric neutrinos are decay products of the particles (mostly pions and kaons) produced as a result of cosmic ray scattering in the Earth's atmosphere. The generated secondary particles decay to $\nu_e$, $\bar{\nu}_e$, $\nu_\mu$ and $\bar{\nu}_\mu$ constituting a significant background to dark matter searches especially for WIMP masses above 100~GeV. In particular, the effect is crucial on the discovery potential of WIMPs with spin-independent cross section of the order of $10^{-48}~\mathrm{cm^2}$. The direct detection dark matter experiments, due to the lack of directional sensitivity, are in principle sensitive to the lowest energy (less than $\sim$100~MeV) Atmospheric neutrinos. For this reason, in our present work Atmospheric neutrinos are considered by employing the low-energy flux coming out of the FLUKA code simulations~\cite{Battistoni:2005pd}.

\subsubsection{Diffuse Supernova Neutrinos}
The weak glow of MeV neutrinos emitted from the total number of core-collapse supernovae, known as the Diffuse Supernova Neutrino Background (DSNB), creates an important source of neutrino background specifically for the WIMPs mass range 10--30~GeV~\cite{Beacom:2010kk}. Despite the appreciably lower flux compared to Solar neutrinos, DSNB neutrino energies are higher than those of the Solar neutrino spectrum. In our simulations, the adopted DSNB distributions (usually of Fermi-Dirac or power-law type) correspond to temperatures 3~MeV for $\nu_e$, 5~MeV for $\bar{\nu}_e$ and 8~MeV for the other neutrino flavours denoted as $\nu_x$ or $\bar{\nu}_x$, $x=\mu,\tau$.  

Figure~\ref{fig:solar-spectrum} shows the unoscillate neutrino flux considered in the present study, illustrating the Solar neutrino spectra of the dominant neutrino sources assuming the high metallicity Standard Solar Model (SSM) as defined in Ref.~\cite{Robertson:2012ib}. Also shown is the low-energy Atmospheric neutrino flux as obtained from the FLUKA simulation~\cite{Battistoni:2005pd} as well as the DSNB spectrum~\cite{Horiuchi:2008jz}. The corresponding neutrino types, maximum energies and fluxes are listed in Table~\ref{table:fluxes}. 

\section{Deformed Shell Model}
\label{sect:DSM}

In the formalism of the WIMP-nucleus or neutrino-nucleus event rates of Sects.~\ref{sect:DM} and~\ref{sect:neutrinos}, 
both for the case of elastic or inelastic interaction channels, the nuclear physics and 
particle physics (SUSY model) parts appear almost completely separated. In the present work our main focus
drops on the nuclear physics aspects which are contained in the nuclear structure factors
discussed in Sect.~\ref{sect:DM}. Special attention is paid on the factors $D_i$ of Eq.(\ref{eq:DM-int2}) 
that depend on the spin structure functions and the nuclear form factors. These quantities
have been calculated using the DSM method~\cite{Sahu:2017czz} (for a comprehensive 
discussion of DSM see Ref.~\cite{ks-book}) given the kinematics and the assumptions 
describing the WIMP particles.

\begin{figure}[t]
\includegraphics[clip, trim=1cm 3cm 6cm 6cm, width=\linewidth]{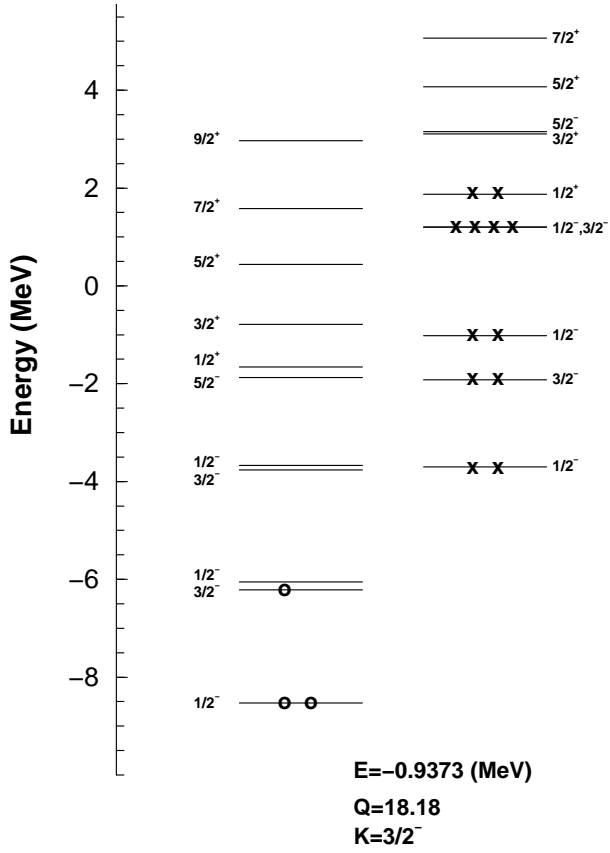}
	\caption{HF single-particle spectra for $^{71}$Ga corresponding to the
	lowest prolate configuration. In the figure, circles represent protons 
	and crosses represent neutrons. The HF energy $E$ in MeV, the mass 
	quadrupole moment $\mathsf{Q}$ in units of the square of the oscillator length
	parameter and the total azimuthal quantum number $K$ are given in the figure.}
\label{fig:HF-spectrum}
\end{figure}

The construction of the many-body wave functions for the initial 
$\vert J^\pi_i\rangle$ and final $\vert J^\pi_f\rangle$ nuclear states in the 
framework of DSM involves performance of the following steps. $(i)$ At first, one 
chooses a model space consisting of a given set of spherical single-particle (sp) 
orbits, sp energies and the appropriate two-body effective interaction matrix elements. 
For $^{71}$Ga and $^{75}$As, the spherical sp orbits are $1p_{3/2}$, $0f_{5/2}$, $1p_{1/2}$ 
and $0g_{9/2}$ with energies 0.0, 2.20, 2.28 and 5.40 MeV and 0.0, 0.78, 1.08 and 3.20 MeV 
respectively, while the assumed effective interaction is the modified Kuo interaction~\cite{ABS}. 
Similarly for $^{127}$I, the sp orbits, their energies and the 
effective interaction are taken from a recent paper~\cite{Coraggio:2017bqn}.
$(ii)$ Assuming axial symmetry and solving the HF single particle 
equations self-consistently, the lowest-energy prolate (or oblate) intrinsic 
state for the nucleus in question is obtained. An example is shown in Fig.~\ref{fig:HF-spectrum}
for $^{71}$Ga.
$(iii)$ The various excited intrinsic states then are obtained by making 
particle-hole ($p$-$h$) excitations over the lowest-energy intrinsic state 
(lowest configuration).
$(iv)$ Then, because the HF intrinsic nuclear states $\vert \chi_K (\eta) 
\rangle$ ($K$ is azimuthal quantum number and $\eta$ distinguishes states 
with same $K$) do not have definite angular momentum, angular momentum projected 
states $\vert \phi^J_{M K} (\mu)\rangle$ are constructed as, 
\begin{equation}
\vert \phi^J_{MK}(\eta)\rangle = \frac{2J+1} {8\pi^2\sqrt{N_{JK}}}\int d\Omega \, D^{J^*}_
{MK}(\Omega)R(\Omega) \vert \chi_K(\eta) \rangle \, .
\end{equation}
In the previous expression,   
$\Omega = (\alpha$, $\beta$, $\gamma$) represents the Euler angles, 
$R(\Omega)$ denotes the known general rotation operator and the Wigner $D$-matrices are defined as $D^{J}_
{MK}(\Omega)= \langle JM \vert R(\Omega)\vert JK \rangle$.
Here, $N_{JK}$ is the normalisation constant which by assuming axial symmetry 
is defined as
\begin{equation}
N_{JK} =\frac{2J+1}{2} \int^\pi_0 d\beta \sin \beta \ d^J_{KK}(\beta)
\langle \chi_K(\eta) \vert e^{-i\beta J_y} \vert \chi_K(\eta)\rangle \, ,
\end{equation}
where the functions $d^J_{KK}(\beta)$ are the diagonal elements of the matrix 
$d^{J}_{MK}(\beta) =\langle J M \vert e^{-i\beta J_y} \vert J K \rangle$.
$(v)$ Finally, the good angular momentum states $\phi^J_{MK}$ are orthonormalised by band mixing calculations and then, in terms of the index $\eta$, it is possible to distinguish between different states having the same angular momentum $J$, 
\begin{equation}
\vert\Phi^J_M(\eta) \rangle \, =\,\sum_{K,\alpha} S^J_{K \eta}(\alpha)\vert 
\phi^J_{M K}(\alpha)\rangle \, .
\end{equation}
Within the DSM method, for the evaluation of the reduced nuclear matrix element 
entering Eqs.(\ref{eq:spin-struct1}) and (\ref{eq:spin-struct2}), we first calculate the single particle matrix elements of the relevant operators $t_\nu^{(l,s)J}$, as
\begin{equation}
\begin{aligned}
\langle n_i & l_i j_i  \vert \vert \hat{t}^{(l,s)J} \vert \vert n_k l_k j_k \rangle =  \\
& \sqrt{(2 j_k + 1) (2 j_i + 1) (2 J + 1) (s+1) (s+2)} \\
& \times  \left\{\begin{array}{ccc}
    {l_i} &  {1/2} &  {j_i}\\
    {l_k} &  {1/2} &  {j_k}\\
    {l}   &  {s}   &  {J}  \end{array} \right \} \langle l_i \vert \vert \sqrt{4 \pi} \, Y^l \vert \vert l_k \rangle \, \langle n_i l_i  \vert j_l (k r) \vert  n_l l_k \rangle \, ,
\end{aligned}
\end{equation}
where $\Bigl\{--\Bigr\}$ is the 9-$j$ symbol.
For more details, the reader is referred to Refs.~\cite{Sahu:2002vc,Sahu:2003xf,Srivastava:2014noa}. It should be noted that in the DSM method one considers an adequate number of intrinsic states in the band mixing calculations. 

DSM calculations are performed in the same spirit as in spherical shell model where one takes a model space and a suitable effective interaction (single particle orbitals, single particle energies and a two-body effective interaction).  This procedure has been found to be quite successful in describing the spectroscopic properties and electromagnetic properties of many nuclei in the mass region A=60--90 and has also been applied to double beta decay nuclear transition matrix elements~\cite{ks-book}. In addition, this model has been used recently in calculating the event rates for dark matter detection~\cite{Sahu:2017czz}. With the proper choice of effective interaction, one will not be considering core excitations. This is a standard prescription in shell model as well as in DSM. To go beyond this, one has to use no-core shell model or DSM with much larger set of single particle orbitals (inclusion of core orbitals), such refinements are planned to be employed in future calculations.

\begin{table*}[t]
\begin{tabular}{ccccccccccccc}
\toprule
Nucleus  & Serial No. &\multicolumn{4}{c}{proton orbits}&\multicolumn{7}{c}{neutron orbits}\\
\midrule
$^{71}$Ga&1& $\pm 1_1$ & $+3_1$  &    &  &$\pm 1_1$  &$\pm 1_2$&$\pm 3_1$ & $\pm 3_2$ &$\pm 1_3$ &$\pm 1^+_1$ & \\
	 &2& $\pm 1_1$ & $+1_2$  &    &  &$\pm 1_1$  &$\pm 1_2$& $\pm 3_1$& $\pm 3_2$ &$\pm 1_3$ & $\pm 1^+_1$\\
	 &3& $\pm 1_1$ & $-3_1$  &    &  &$\pm 1_1$  &$\pm 1_2$& $\pm 3_1$& $\pm 3_2$ & $\pm 1_3$& $+1^+$  & $+3^+$\\
	 &4& $\pm 1_1$ & $+3_1$  &    &  &$\pm 1_1$  &$\pm 1_2$& $\pm 3_1$& $\pm 3_2$ &$\pm 1_3$ & $\pm 5_1$& \\
	 & &           &         &    &  &           &         &          &           &          &          &     \\
$^{73}$Ge&1& $\pm 1_1$ &$\pm 1_2$&    &  &$\pm 1_1$  &$\pm 1_2$&$\pm 3_1$ &$\pm 3_2$ & $\pm 1_3$ &$\pm 1^+_1$& $+3_1^+$ \\
         &2& $\pm 1_1$ &$\pm 1_2$&    &  &$\pm 1_1$  &$\pm 1_2$&$\pm 3_1$ &$\pm 3_2$ & $\pm 1_3$ &$\pm 3_1^+$& $+1_1^+$ \\
	 &3& $\pm 1_1$ &$\pm 3_1$&    &  &$\pm 1_1$  &$\pm 1_2$&$\pm 3_1$ &$\pm 3_2$ & $\pm 1_3$ &$\pm 1_1^+$& $+3_1^+$ \\
	 & &           &         &    &  &           &         &          &           &          &          &     \\
$^{75}$As&1&$\pm 1_1$ &$\pm 1_2$&$+3_1$& &$\pm 1_1$  &$\pm 1_2$&$\pm 3_1$ &$\pm 1_1^+$&$\pm 3_1^+$&$\pm 3_2$& $\pm 1_3$ \\
	 &2& $\pm 1_1$&$\pm 3_1$&$+1_2$& &$\pm 1_1$  &$\pm 1_2$&$\pm 3_1$ &$\pm 1_1^+$&$\pm 3_1^+$&$\pm 3_2$&$\pm 1_3$   \\
	 &3& $\pm 1_1$&$\pm 1_2$&$+3_1$& &$\pm 1_1$  &$\pm 1_2$&$\pm 3_1$ &$\pm 1_1^+$&$\pm 3_1^+$&$\pm 3_2$&$\pm 5_1^+$\\
	 &4& $\pm 1_1$&$\pm 3_1$&$+1_3$& &$\pm 1_1$  &$\pm 1_2$&$\pm 3_1$ &$\pm 1_1^+$&$\pm 3_1^+$&$\pm 3_2$&$\pm 5_1^+$ \\
	 &5&$\pm 1_1$ &$\pm 1_2$&$+3_1$& &$\pm 1_1$  &$\pm 1_2$&$\pm 3_1$ &$\pm 1_1^+$&$\pm 3_1^+$&$\pm 3_2$  &$\pm 5_1$ \\
	 &6&$\pm 1_1$ &$\pm 3_1$&$+1_2$& &$\pm 1_1$  &$\pm 1_2$&$\pm 3_1$ &$\pm 1_1^+$&$\pm 3_1^+$&$\pm 3_2$  &$\pm 5_1$ \\
	 & &           &         &  & &           &         &          &           &          &          &     \\
$^{127}$I&1&$\pm 7_1^+$&$+5_1^+$ &  & &$\pm 7_1^+$&$\pm 5_1^+$&$\pm 3_1^+$& $\pm 11_1$&$\pm 1_1^+$&$\pm 5_2^+$& $\pm 9_1$\\
	 & &           &         &  & &$\pm 3_2^+$ &$\pm 1_2^+$&$\pm 7_1$  &$\pm 5_1$  &$\pm 3_1$ & &\\
	 &2&$\pm 7_1^+$& $+5_1^+$&  & &$\pm 7_1^+$&$\pm 5_1^+$&$\pm 3_1^+$& $\pm 11_1$&$\pm 1_1^+$&$\pm 5_2^+$& $\pm 9_1$\\
	 & &           &         &  & &$\pm 3_2^+$&$\pm 1_2^+$ &$\pm 7_1$  &$\pm 5_1$  &$\pm 3_3^+$& &  \\
	 &3&$\pm 7_1^+$& $+3_1^+$&  & &$\pm 7_1^+$&$\pm 5_1^+$&$\pm 3_1^+$& $\pm 11_1$&$\pm 1_1^+$&$\pm 5_2^+$& $\pm 9_1$\\
	 & &           &         &  & &$\pm 3_2^+$&$\pm 1_2^+$&$\pm 7_1$  &$\pm 5_1$  &$\pm 3_1$  \\
	 &4&$\pm 7_1^+$& $+3_1^+$&  & &$\pm 7_1^+$&$\pm 5_1^+$&$\pm 3_1^+$& $\pm 11_1$&$\pm 1_1^+$&$\pm 5_2^+$& $\pm 9_1$\\
	 & &           &         &  & &$\pm 3_2^+$&$\pm 1_2^+$&$\pm 7_1$  &$\pm 5_1$  &$\pm 3_3^+$  \\
         &5&$\pm 7_1^+$& $+1_1^+$&  & &$\pm 7_1^+$&$\pm 5_1^+$&$\pm 3_1^+$& $\pm 11_1$&$\pm 1_1^+$&$\pm 5_2^+$& $\pm 9_1$\\
	 & &           &         &  & &$\pm 3_2^+$&$\pm 1_2^+$&$\pm 7_1$  &$\pm 5_1$  &$\pm 3_1$  \\
	 &6&$\pm 7_1^+$& $+1_1^+$&  & &$\pm 7_1^+$&$\pm 5_1^+$&$\pm 3_1^+$& $\pm 11_1$&$\pm 1_1^+$&$\pm 5_2^+$& $\pm 9_1$\\
	 & &           &         &  & &$\pm 3_2^+$&$\pm 1_2^+$&$\pm 7_1$  &$\pm 5_1$  &$\pm 3_3^+$  \\
\bottomrule
\end{tabular}
\caption{$2k$ values of the occupied proton and neutron single particle deformed orbits of the HF intrinsic states used in the calculation for each nucleus. The second column gives the serial no. of the HF intrinsic states used.  All the $2k$ values are of negative parity unless explicitly shown. The $(+)$ , $(-)$ or $(\pm)$ sign before the $2k$ values implies that either the time-like,  time-reversed, or both orbits are occupied.  In columns 3 and 4, $3_1$ means the first $3/2^-$ HF deformed sp orbit, $3_2$ means the second $3/2^-$ deformed HF orbit and so on (see also Fig.~\ref{fig:HF-spectrum}). Detailed information regarding the structure each of the deformed HF sp orbits,  their energies as well as the parentage of each of the HF intrinsic state in the $\Phi^J$ states (e.g. the linear combination of $\phi^J_{MK}$ obtained in the band mixing diagonalisation)  can be obtained from the authors.}
\label{table:orbits}
\end{table*}

We note that the many body nuclear calculations performed take into account in the usual way the inert core orbits (completely filled by the protons and neutrons) and the extra-core nucleons moving in the assumed model space under the influence of an effective interaction. The explicit $2k$ values of the occupied nucleon single particle deformed orbits of the HF intrinsic states considered in our calculations are listed in Table~\ref{table:orbits}.

\section{Results and Discussion}
\label{sect:results}

\subsection{Nuclear physics aspects}

\begin{table*} 
\begin{tabular}{ccccccccccc}
\toprule
Nucleus & $A$ & $Z$ & $J^\pi$   & $<l_p>$ & $<S_p>$ & $<l_n>$ & $<S_n>$ & $\mu$ (nm) & Exp & $b \, \, [\mathrm{fm^{-1}}]$ \\
\midrule
Ga & 71 & 31 & $3/2^-$ & 0.863 & 0.257 & 0.369 & 0.011 & 2.259 & 2.562 & 1.90 \\
Ge & 73 & 32 & $9/2^+$ & 0.581 & $-0.001$ & 3.558 & 0.362 & $-0.811$ & $-0.879$ & 1.91 \\
As & 75 & 33 & $(3/2^-)_1$ &  0.667  &   0.164  &  0.626  &  0.042  &  1.422  &  1.439 & 1.92 \\
I & 127 & 53 & $5/2^+$ & 2.395 & $-0.211$ & 0.313 & 2.343 & 1.207 & 2.813 & 2.09\\   
\bottomrule
\end{tabular}
\caption{List of potential Dark Matter detectors considered in the present study. The calculated magnetic moments for the ground states of $^{71}$Ga, $^{73}$Ge, $^{75}$As and $^{127}$I, are shown. The results involve the bare gyromagnetic ratios and experimental data are from~\cite{nndc}. The ground state $J^\pi$ and the harmonic oscillator size $b$ are also shown.}
\label{tab-1}
\end{table*}
%
\begin{figure}[t]
\includegraphics[clip, trim=1cm 4cm 6cm 6cm, width=\linewidth]{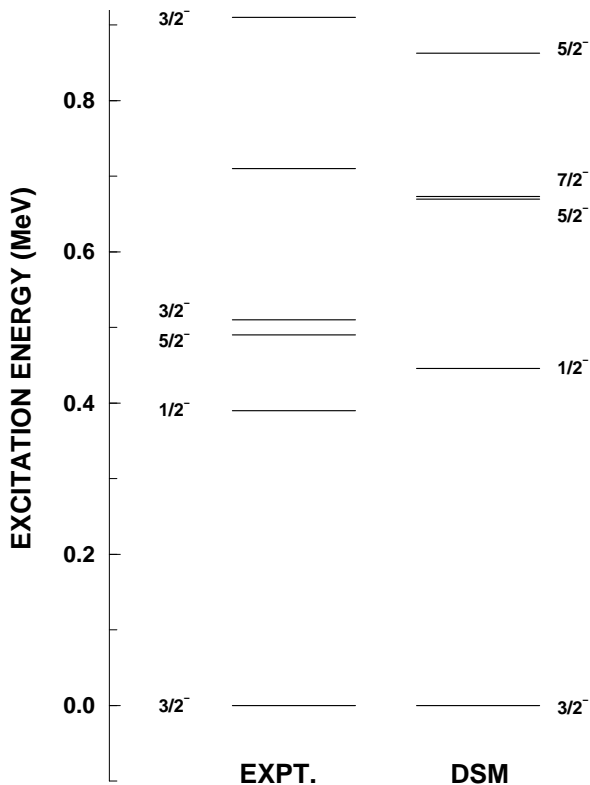}
\caption{Comparison of deformed shell model results with experimental data for $^{71}$Ga for low-lying states. The experimental values are taken from~\cite{nndc}.}
\label{fig:DSM-spectrum}
\end{figure}

\begin{figure}[t]
\centering
\includegraphics[width=\linewidth]{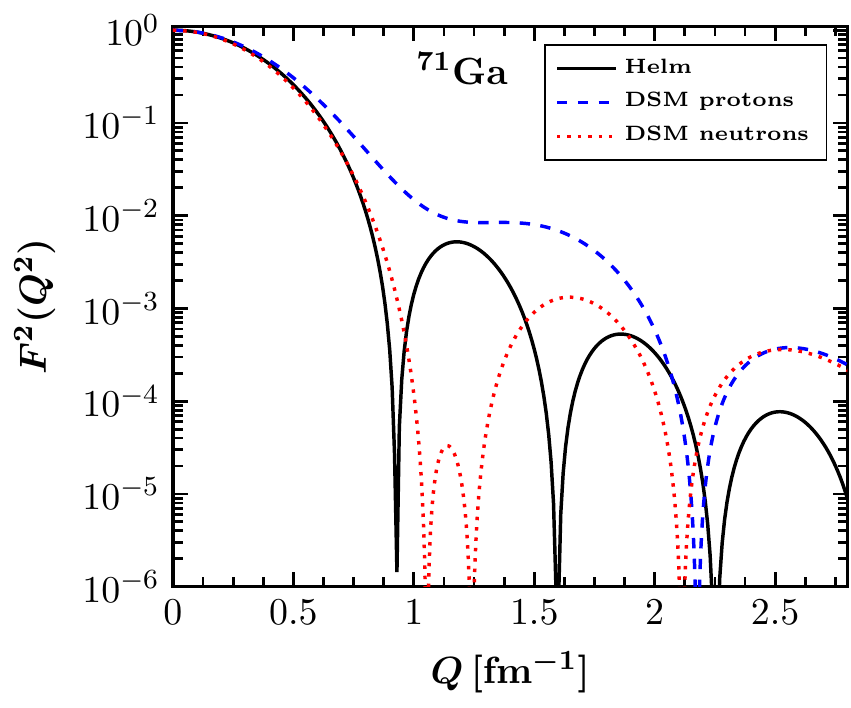}\\
\includegraphics[width=\linewidth]{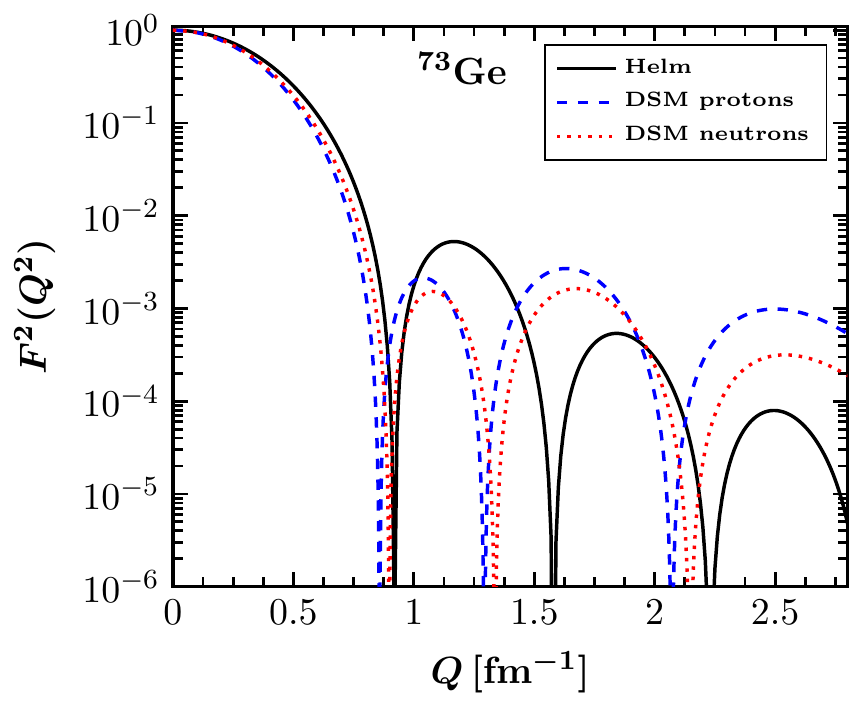}
\caption{Comparison of the DSM nuclear form factors of the $^{71}$Ga and $^{73}$Ge isotopes, obtained in the present work with the corresponding effective Helm form factors.}
\label{fig:formfac1}
\end{figure}
\begin{figure}[t]
\centering
\includegraphics[width= \linewidth]{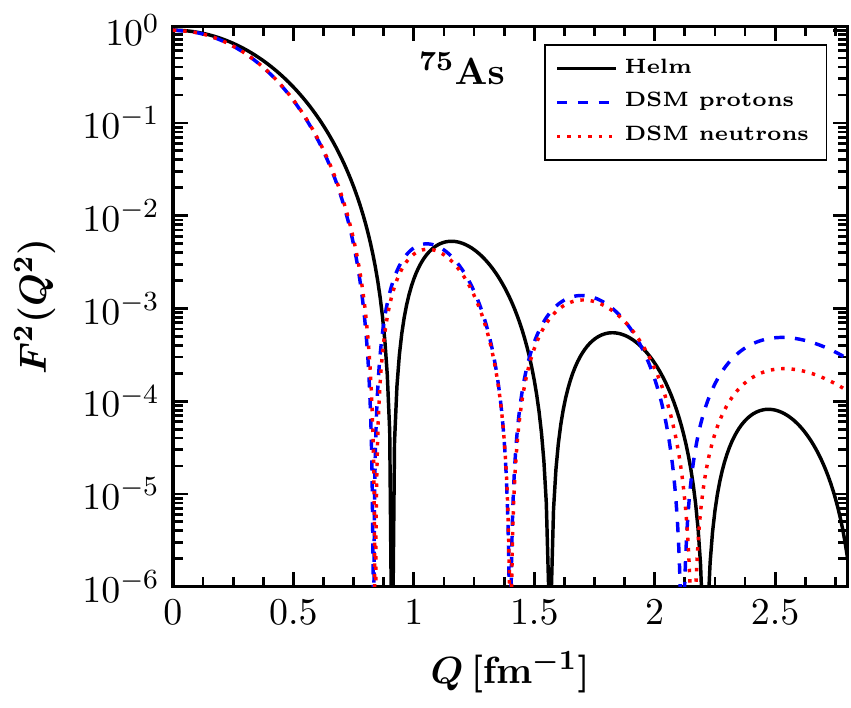}
\includegraphics[width= \linewidth]{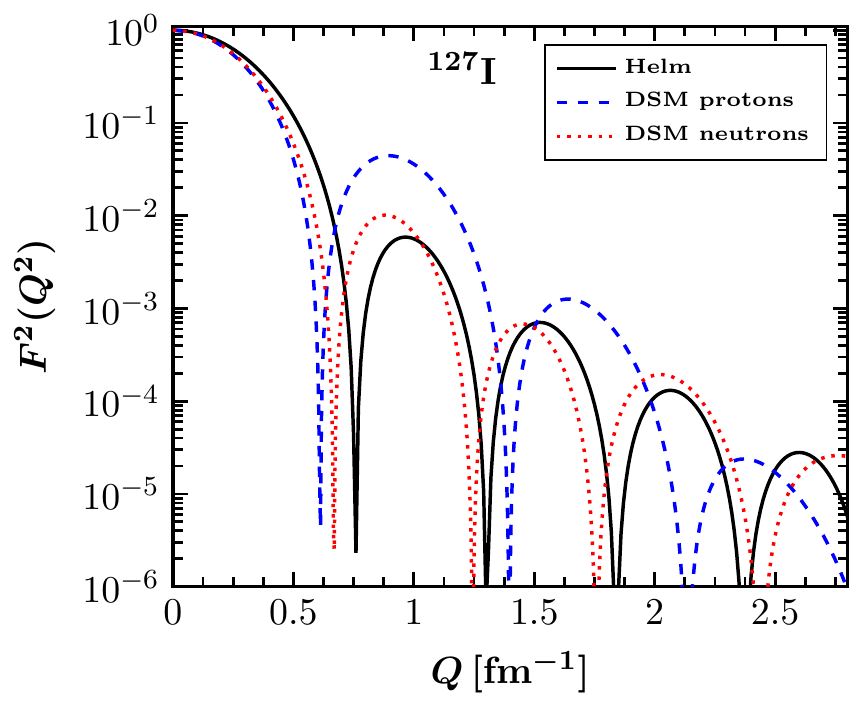}
\caption{Same as in Fig.~\ref{fig:formfac1} but for the $^{75}$As and $^{127}$ I isotopes.}
\label{fig:formfac2}
\end{figure}

\begin{figure}[ht]
\centering
\includegraphics[width=\linewidth]{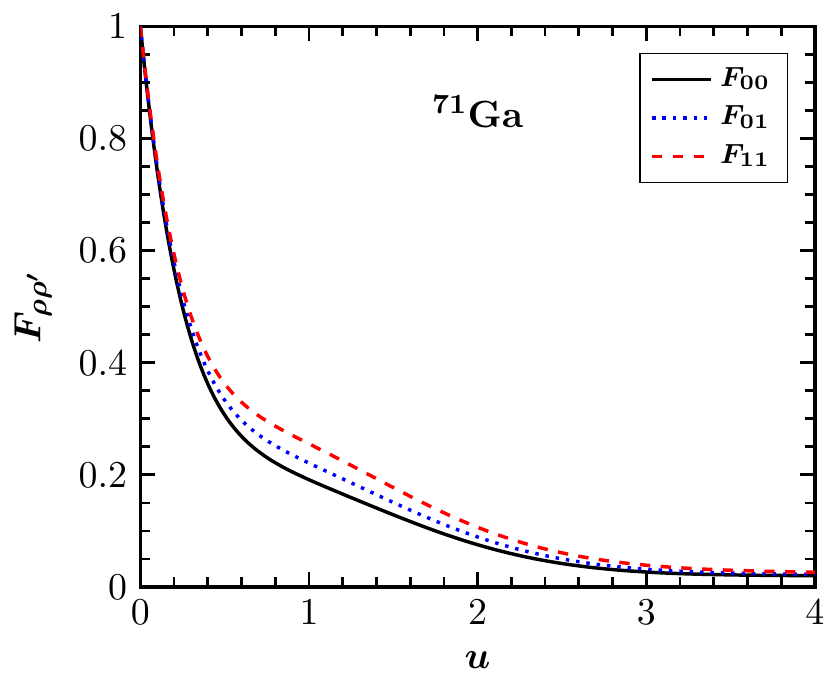}
\caption{Normalised spin structure functions of $^{71}$Ga for the ground state.}
\label{fig:spinstruct-Ga71-elastic}
\end{figure}
To maximise the significance of our WIMP-nucleus and neutrino-floor calculations, the reliability of the obtained nuclear wave functions is tested by comparing the extracted energy  level spectrum and magnetic moments with available experimental  data. The consistency of this method obtained for $^{73}$Ge has been already presented in Ref.~\cite{Sahu:2017czz}. Furthermore, in the DSM calculations for $^{71}$Ga and $^{75}$As, we restrict ourselves to prolate solutions only, since the oblate solution does not reproduce the energy spectra and electromagnetic properties of these nuclei. It also does not mix with the prolate solution. Hence we neglect the oblate solutions in the calculations. For each of these nuclei, we consider only four intrinsic prolate states which should be sufficient to explain the systematics of the ground state and close lying excited state.  Due to size restrictions, in Fig.~\ref{fig:DSM-spectrum} we illustrate only the calculated spectrum for $^{71}$Ga.

For $^{75}$As, the ground state is $3/2^-$ and there are also two $1/2^-$ and $3/2^-$ levels around 0.12 MeV. In addition there is a collective band consisting of $5/2^-$, $9/2^-$, ${13/2^-}$ ${17/2^-}$ levels at 0.279, 1.095, 2.150 and 3.091~MeV, respectively.  All these levels are well reproduced by the DSM method. Turning to the $^{127}$I spectrum, there are four observed collective bands with band heads $5/2^+$, $(7/2^+)_{1,2}$ and $9/2^+$. There are evidences suggesting that low-lying states in $^{127}$I have oblate deformation~\cite{Ding:2012zzb}. Hence for this nucleus, we consider only oblate  configurations and take the six lowest oblate intrinsic states in the band mixing calculation. These intrinsic states are found to provide adequate description of the energy spectrum and electromagnetic  properties for this nucleus.  The calculations for  this nucleus utilise a new effective interaction developed by an Italian group very recently~\cite{Coraggio:2017bqn}. The new effective interaction is seen to reproduce well the $^{127}$I spectrum; details will be presented elsewhere. 

We thus conclude that concerning the evaluation of the WIMP-nucleus and CE$\nu$NS event rates we are interested in this work, the required ground state wave functions obtained through the DSM method are reliable and the intrinsic states used in the subsequent analysis are considered sufficient. From the perspective of nuclear physics, spin contributions constitute significant ingredients in the evaluation of WIMP-nucleus event rates. For this reason, the first stage of our work involves the calculation of the magnetic moment, which is decomposed into an orbital and spin part. The  relevant results for the proton and neutron contributions to the orbital and spin parts concerning the ground states of the four nuclear isotopes  studied in this paper are given in Table~\ref{tab-1}. A comparison between the obtained magnetic moments and the respective experimental data is also provided. Despite the fact that, these calculations adopt bare values of $g$-factors neglecting quenching effects, the obtained DSM result of the ground state magnetic moments are consistent with the experimental values. 

Having successfully reproduced the energy spectrum and the magnetic moments within the context of the DSM wave functions, we evaluate important nuclear physics inputs entering the WIMP-nucleus and CE$\nu$NS cross sections. Figures~\ref{fig:formfac1} and \ref{fig:formfac2} present a comparison between the DSM nuclear form factors and the effective Helm-type ones employed in various similar studies, where as can be seen, the DSM results differ from the Helm-type ones. The behaviour of the proton form factor for $^{71}$Ga is found to be different from those of the other nuclei and this may be due to the nearby proton shell closure and the neutron sub-shell closure. Calculations with several different effective interactions are under way to rule out the possibility of any deficiency of the effective two-body interaction used. We furthermore, illustrate the spin structure functions of WIMP-$^{71}$Ga elastic scattering calculated using Eqs.(\ref{eq:spin-struct1}) and (\ref{eq:spin-struct2}). The variation of $F_{00}$, $F_{01}$ and $F_{11}$ with respect to the parameter $u$ is shown in Fig.~\ref{fig:spinstruct-Ga71-elastic}, while similar results are obtained for $^{75}$As and $^{127}$I (for the $^{73}$Ge case see Ref.~\cite{Sahu:2017czz}).

The consistency of our nuclear physics DSM calculations has been extensively explored in this work and compared with existing experimental data (see Figs.~\ref{fig:HF-spectrum}, \ref{fig:DSM-spectrum} and Table~\ref{tab-1}) making the considered form factors reliable. Specifically we have tested the reliability of this model to describe nuclear-structure properties such as excitation spectra and nuclear magnetic moments. We mention that DSM has been tested in the past in many nuclei in the $A$=60--90 region~\cite{ks-book} (see above).

\subsection{WIMP-nucleus rates and the neutrino-floor}
\begin{figure}[t]
\centering
\includegraphics[width=\linewidth]{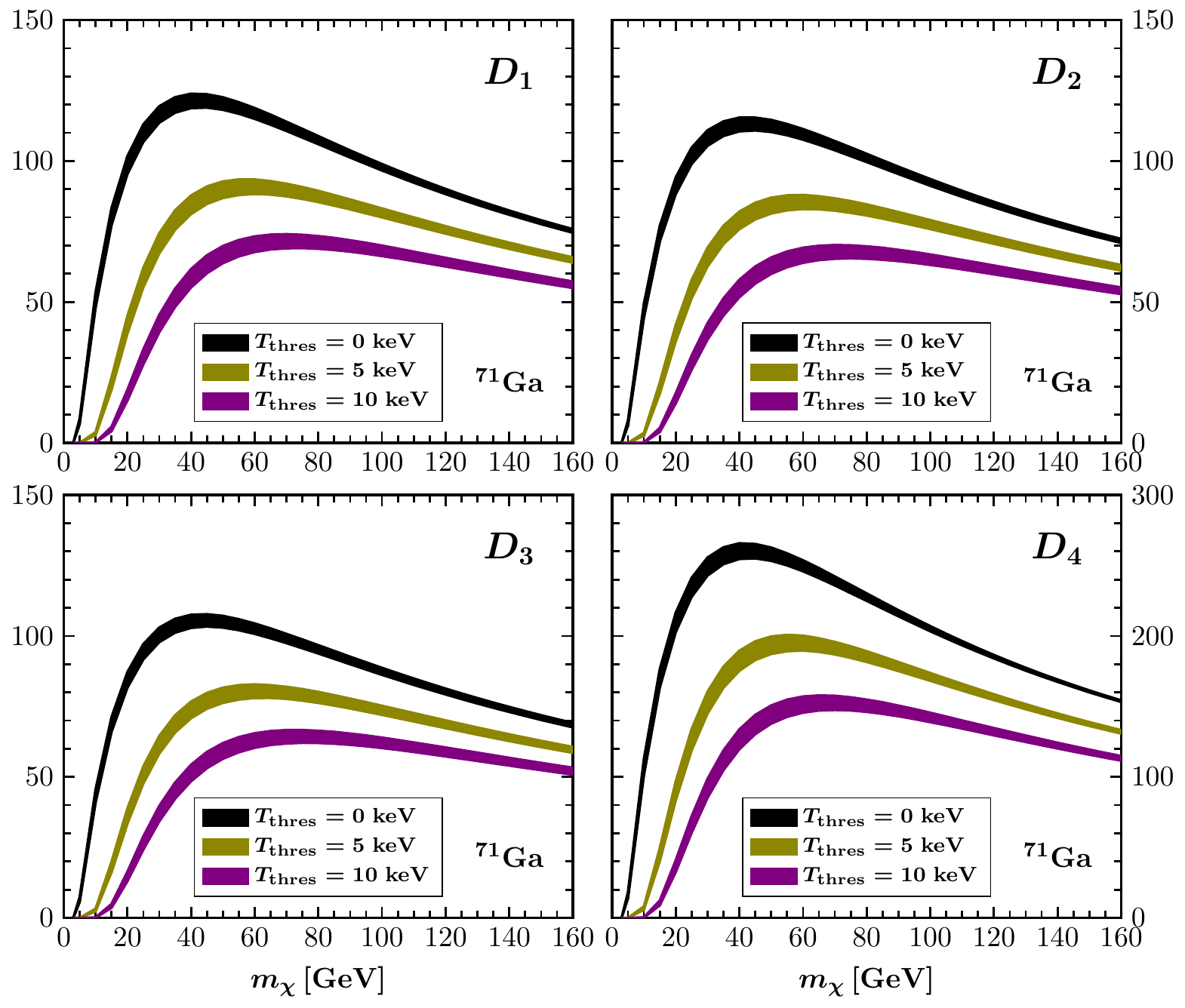}
\caption{ Nuclear structure coefficients $D_i$ for $^{71}$Ga plotted as a function of the WIMP mass. The graphs are plotted for three values of the detector threshold 	0, 5, 10~keV. The thickness of the graphs represent annual modulation.}
\label{fig:spin-struct-Ga71-elastic}
\end{figure}

The WIMP-nucleus event rates and the neutrino-floor due to neutrino-nucleus scattering are calculated for a set of interesting nuclear targets such as $^{71}$Ga, $^{73}$Ge, $^{75}$As and $^{127}$I (see Table~\ref{tab-1}). In evaluating the neutrino-induced backgrounds, we consider only the dominant CE$\nu$NS channel, since neutrino-electron events are expected to produce less events by about one order of magnitude~\cite{Billard:2013qya}.  For the case of a $^{71}$Ga target, in Fig.~\ref{fig:spin-struct-Ga71-elastic} we provide the coefficients $D_i$ associated to the spin dependent and coherent interactions given in Eq.(\ref{eq:DM-int2}) as functions of the WIMP mass $m_\chi$ by assuming three typical values of the detector threshold energy $T_N$= 0, 5, 10~keV. For the special case of $T_N$ = 0, all plots peak at $m_\chi\sim$35~GeV, while for higher threshold energies $D_i$ are shifted towards higher values of the WIMP mass. The calculations take also into account the annual modulation which is represented by the curve thickness. As can be seen from the figure, the modulation signal varies with respect to the WIMP mass, being larger for $m_\chi \leq 50$~GeV while its magnitude is slightly different for the spin dependent and coherent channels. 

\begin{figure}[t]
\centering
\includegraphics[width=\linewidth]{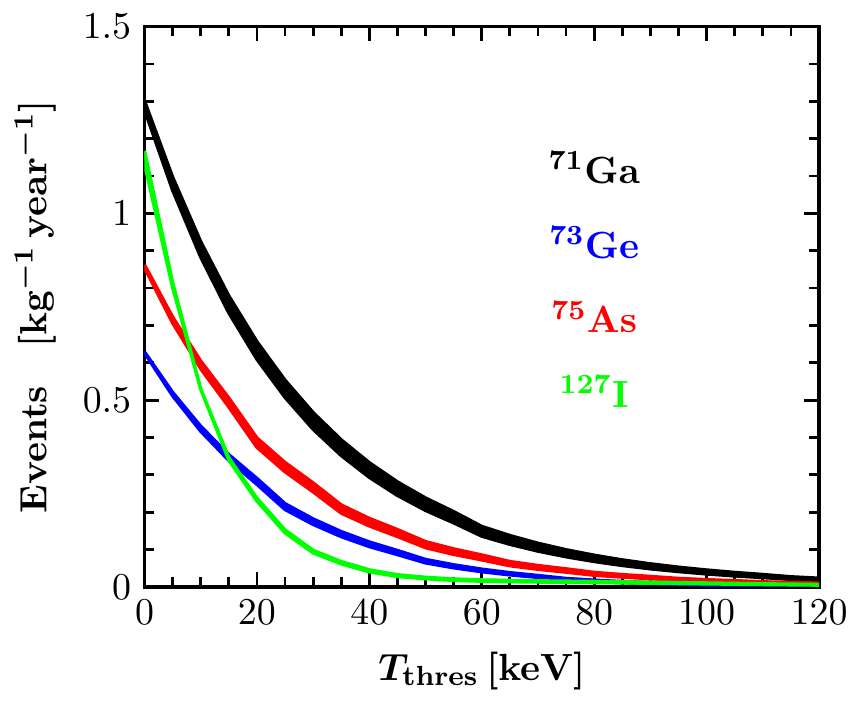}
\caption{The WIMP event rates for $^{71}$Ga, $^{73}$Ge, $^{75}$As and $^{127}$I detectors in units of $\mathrm{kg^{-1} \, year^{-1}}$ as a function of the detector threshold $T_N$. The nuclear threshold $T_N$ energy through the limit of the integration in Eq.(\ref{eq:DM-int2}). The thickness of the curve represents the annual modulation which decreases with increasing nuclear mass. }
\label{fig:DM-events}
\end{figure}

Proceeding further, in Fig.~\ref{fig:DM-events} we evaluate the expected event rates for the four target nuclei assuming elastic WIMP scattering for WIMP candidates with mass $m_\chi=110$~GeV, by adopting the nucleonic current parameters $f^0_A=3.55 \times 10^{-2}$, $f^1_A=5.31 \times 10^{-2}$, $f^0_S=8.02 \times 10^{-4}$ and $f^1_S=-0.15\times f^0_S$. As in the previous discussion, the thickness of the graph accounts for the annual modulation. We find that there is a strong dependence of the event rate on the studied nuclear isotope. Again the modulation is found to decrease for heavier mass. Among the four studied nuclei, we come out with a larger event rate for the case of a $^{71}$Ga nuclear detector, since $D_1$, $D_2$, $D_3$ are all positive and have similar values. For $^{73}$Ge, $D_2$ is negative and its magnitude is comparable to $D_1$ and $D_3$, while for $^{75}$As, $D_3$ is positive but small, and finally for $^{127}$I, $D_2$ and $D_3$ are relatively smaller and $D_1$ is large. The coherent contribution $D_4$ has more or less similar values for all nuclei considered.
\begin{figure*}[ht!]
\centering
\includegraphics[width= 0.8\linewidth]{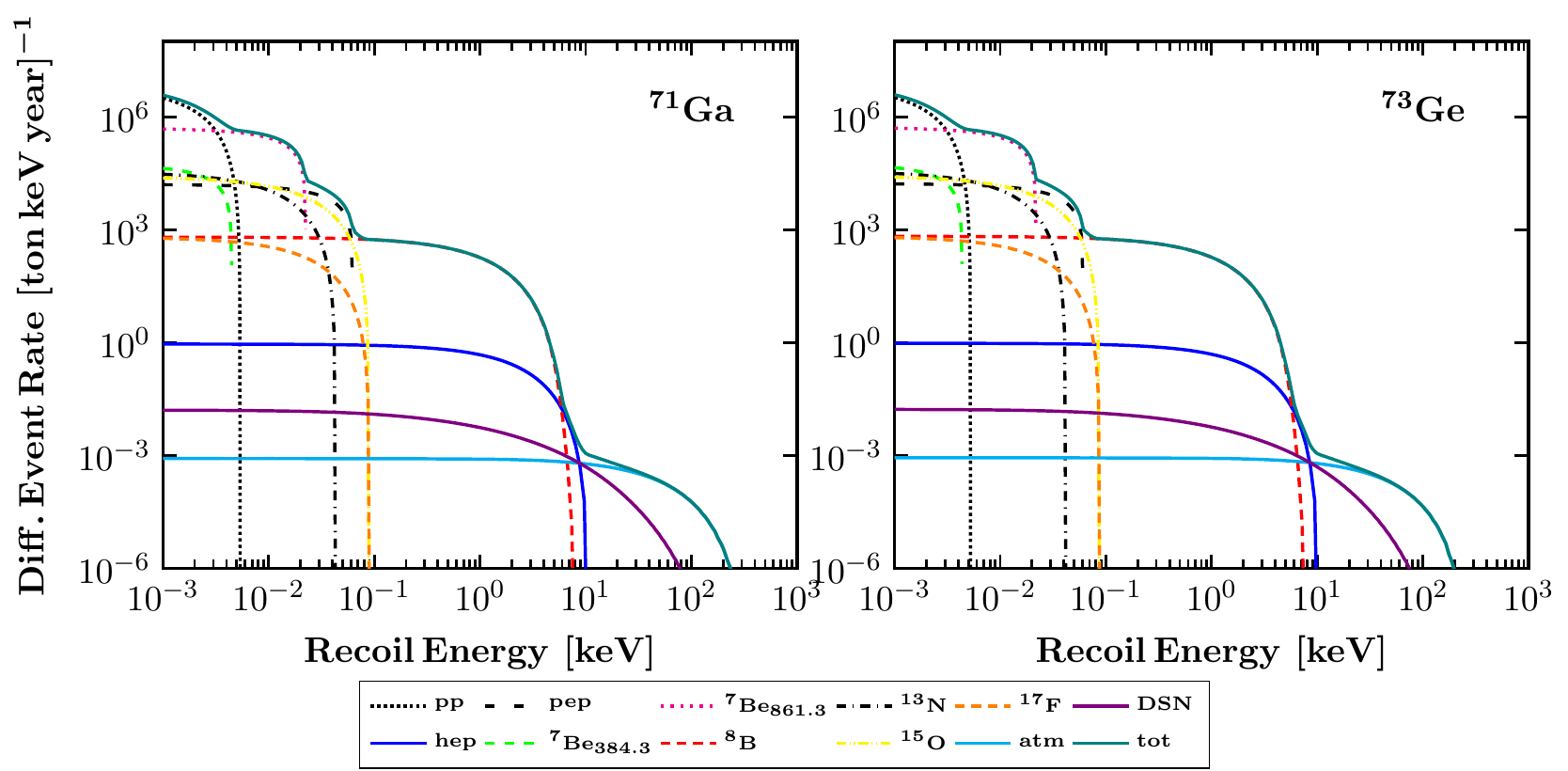} 
\includegraphics[width= 0.8\linewidth]{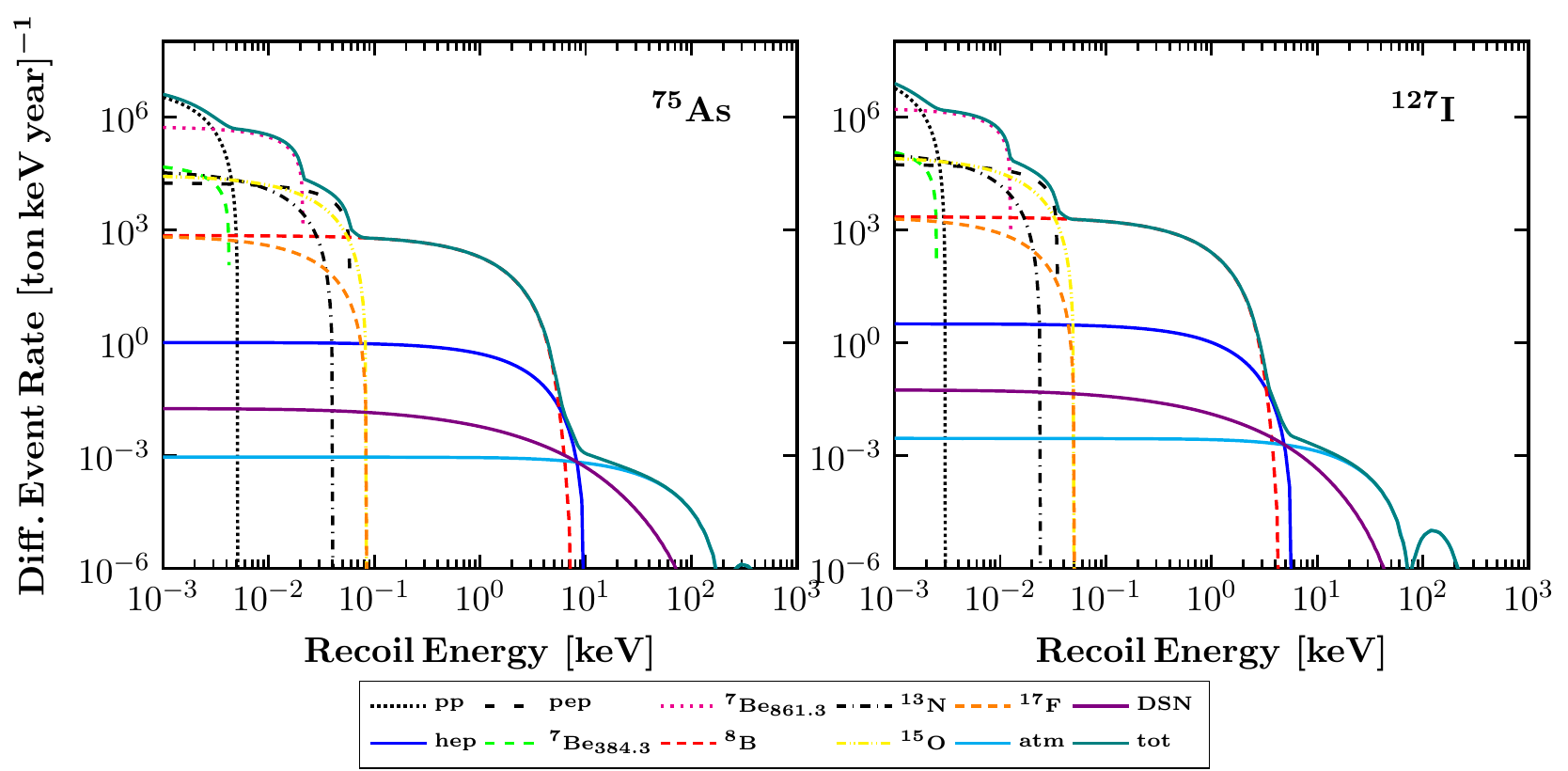} 
\caption{Differential event rate of the neutrino-floor assuming $^{71}$Ga, $^{73}$Ge, $^{75}$As and $^{127}$I as cold dark matter detectors. The individual components coming from the Solar, Atmospheric and DSNB flux are also shown.}
\label{fig:solar_event-rate}
\end{figure*}
\begin{figure*}[ht!]
\centering
\includegraphics[width=0.8 \textwidth]{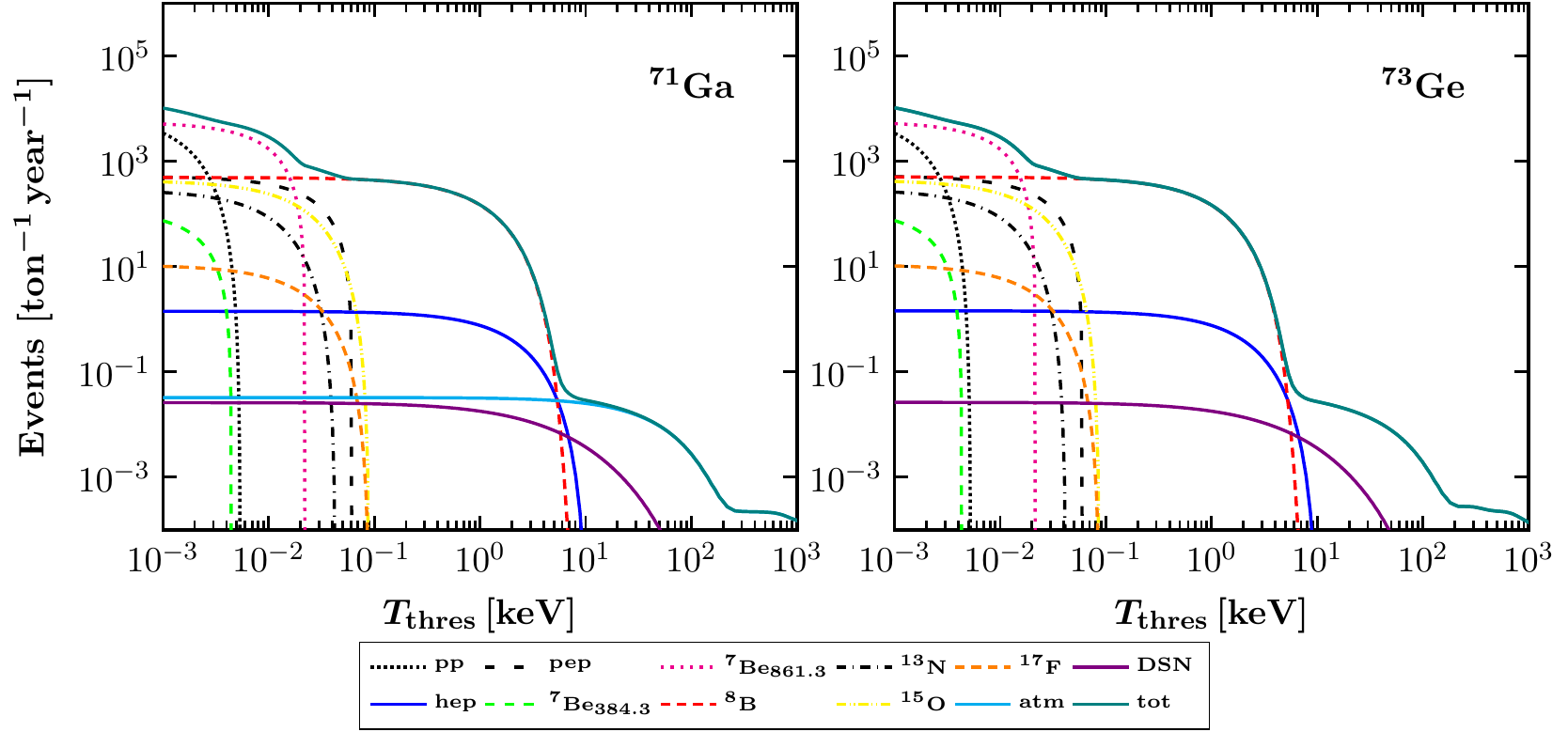} 
\includegraphics[width=0.8 \textwidth]{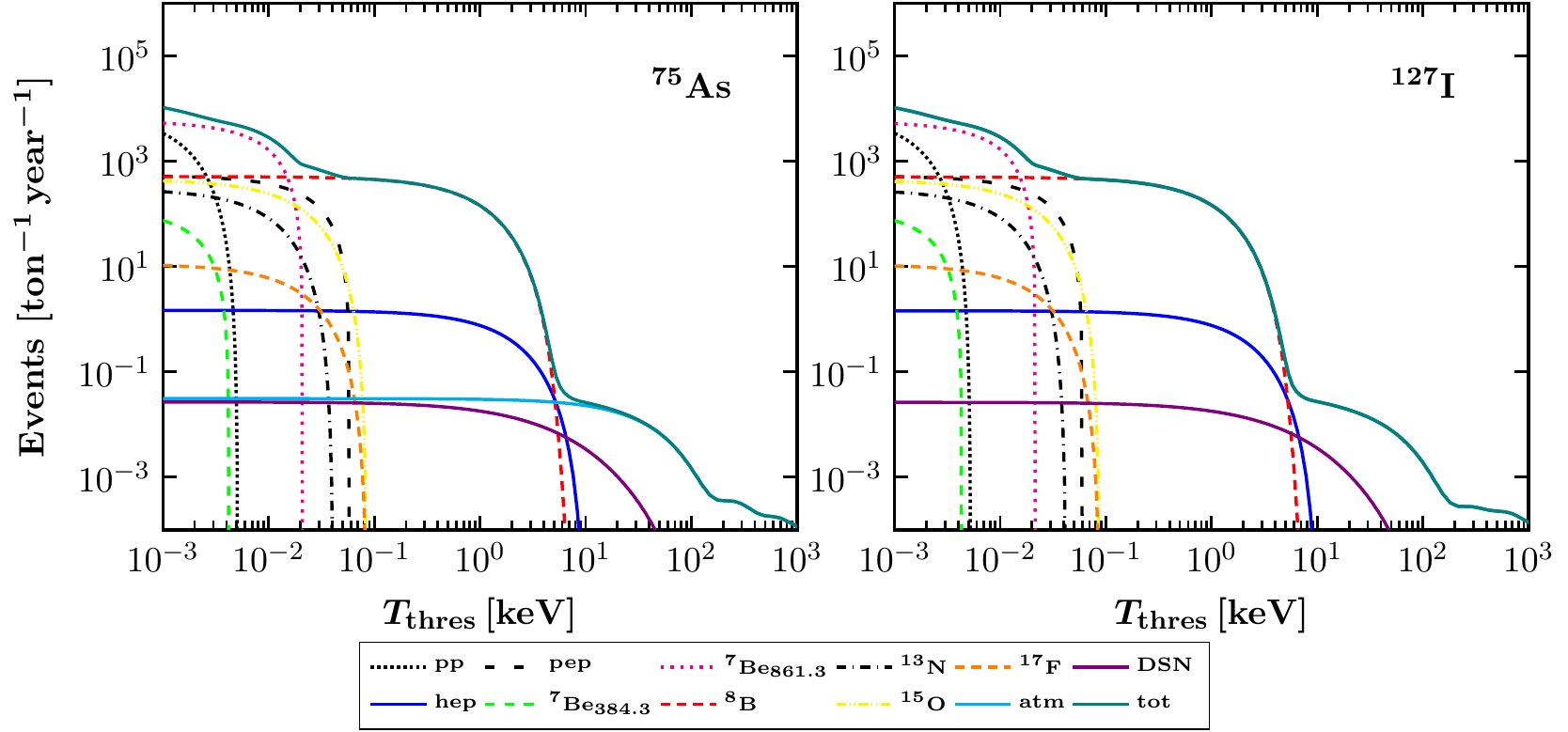} 
\caption{Same as in Fig.~\ref{fig:solar_event-rate} but for the number of events above the detector threshold.}
\label{fig:solar_events}
\end{figure*}

For each component of the Solar, Atmospheric and DSNB neutrino distributions we calculate the expected neutrino-floor due to CE$\nu$NS, by considering the target nuclei presented in Table~\ref{tab-1}. In our calculations, we neglect possible recoil events arising from Geoneutrinos as they are expected to be at least one order of magnitude less that the aforementioned neutrino sources (see e.g. Refs.~\cite{Monroe:2007xp,Gelmini:2018ogy}). In order to make a quantitative estimate of the neutrino-floor, here we do not consider neutrino oscillations and we assume that CE$\nu$NS is a flavour blind process in the SM. The differential event rate due to CE$\nu$NS, for the various dark matter detectors considered in the present study, is presented in Fig.~\ref{fig:solar_event-rate}. It can be noticed that, the neutrino-background is dominated by Solar neutrinos at very low recoil energies. We stress that, for the typical keV-recoil thresholds of the current direct detection dark matter experiments only the $hep$ and $^8$B sources constitute a possibly detectable background. From our results we conclude that, for recoil energies above about 10~keV, Atmospheric neutrinos dominate the neutrino background event rates, having a tiny contribution coming from the DSNB spectrum. 

The number of expected background events due to CE$\nu$NS for each component of the Solar, Atmospheric and DSNB neutrino fluxes is illustrated in Fig.~\ref{fig:solar_events}. Similarly to the differential event case, at low energies the neutrino background is dominated by the Solar neutrino spectrum with, the dominant components being the $hep$ and $^8$B neutrino sources. The results imply that future multi-ton scale detectors with sub-keV sensitivities may be also sensitive to $^7$Be and $pp$ neutrinos. We comment however that, such sensitivities will be further limited due to the quenching effect of the nuclear recoil spectrum which is not taken into account here. Moreover, it is worth mentioning that neutrino-induced and WIMP-nucleus scattering processes provide similar recoil spectra, e.g. the recoil spectrum of $^8$B  neutrinos may mimic that of a WIMP with mass 6~GeV (100~MeV)~\cite{OHare:2016pjy}.

\begin{figure}[t]
\centering
\includegraphics[width= \linewidth]{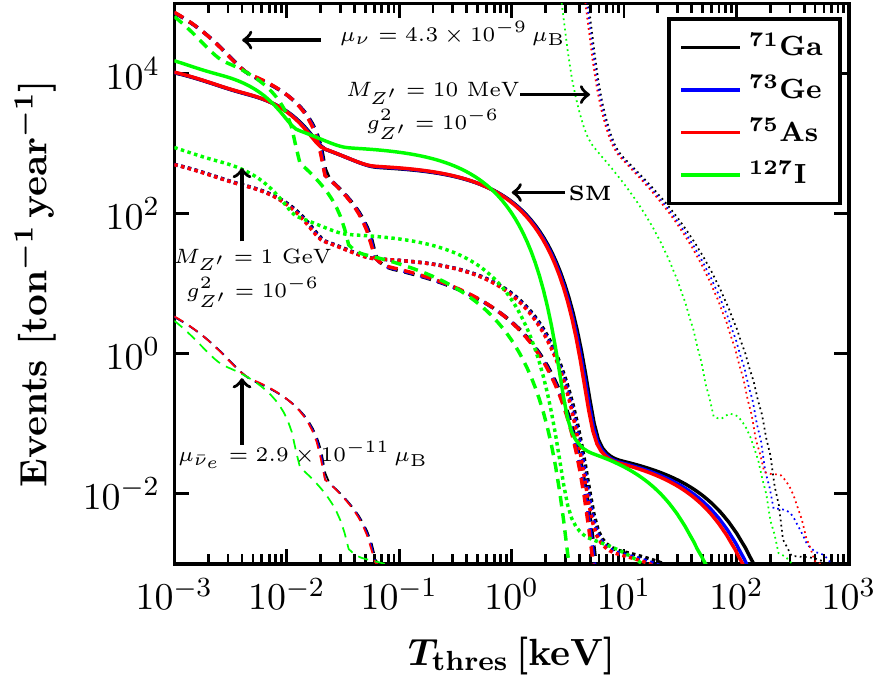} 
\caption{The neutrino-floor for various interaction channels. Solid, dashed and dotted lines correspond to SM, EM and $Z^\prime$ contributions respectively.}
\label{fig:mag_events}
\end{figure}

At this point, we consider additional interactions in the context of new physics beyond the SM that may enhance the CE$\nu$NS rate at a direct detection dark matter experiment. Specifically we study the impact of neutrino EM properties as well as the impact of new interactions due to a $Z^\prime$ mediator, on the neutrino floor. In our calculations we assume the existence of a neutrino magnetic moment $\mu_\nu = 4.3 \times 10^{-9} \mu_B$, extracted from CE$\nu$NS data in Ref.~\cite{Kosmas:2017tsq} as well as the corresponding limit from $\bar{\nu}_e-e^-$ scattering data of the GEMMA experiment, e.g. $\mu_{\bar{\nu}_e} = 2.9 \times 10^{-11} \mu_B$~\cite{Beda:2012zz}. Regarding the $Z^\prime$ interaction we consider typical values such as $M_Z^\prime=10~\mathrm{MeV}$, $g_{Z^\prime}^2=10^{-6}$ and $M_Z^\prime=1~\mathrm{GeV}$, $g_{Z^\prime}^2=10^{-6}$~\cite{Billard:2018jnl}. Following Ref.~\cite{Liao:2017uzy}, by assuming universal couplings, our calculations involve the product of neutrino and quark $Z^\prime$ couplings defined as (for a comprehensive study involving the flavour dependence of the $Z^\prime$ couplings the reader is refereed to Ref.~\cite{Abdullah:2018ykz})
\begin{equation}
g_{Z^\prime}^2 = \frac{g_{Z^\prime}^{\nu V} \mathcal{Q}_{Z^\prime}}{3 A} \, .
\end{equation}
The corresponding results are presented in Fig.~\ref{fig:mag_events}, indicating that such new physics phenomena may constitute a crucial source of background even for multi-ton scale detectors with sub-keV capabilities. We stress however, that the latter conclusion depends largely on the assumed parameters, which currently are unknown.

\begin{figure}[ht!]
\centering
\includegraphics[width= \linewidth]{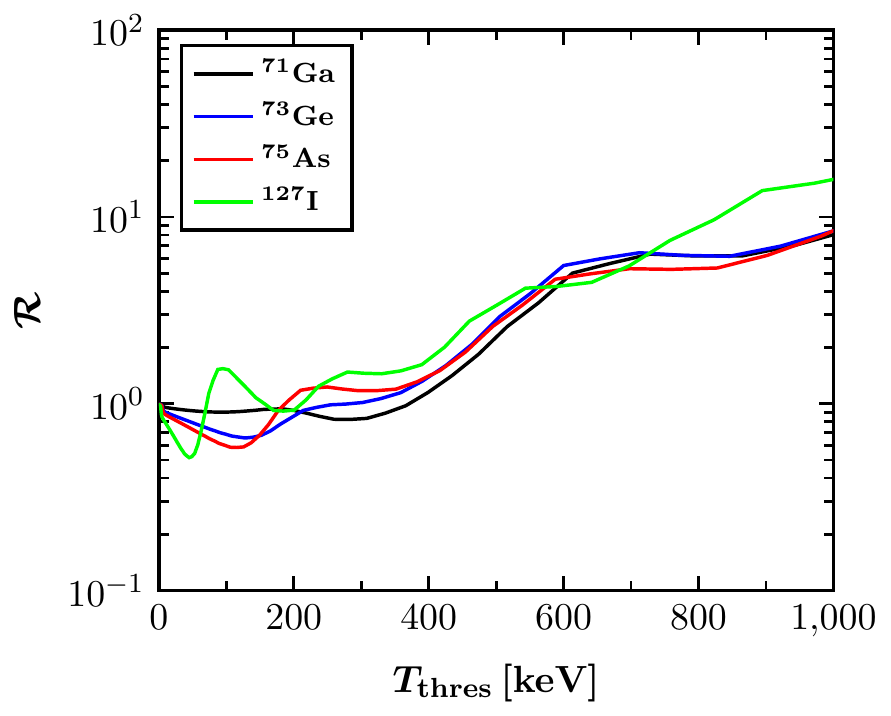} 
\caption{The ratio $\mathcal{R}$ as a function if the detector threshold.}
\label{fig:DSM_over_Helm}
\end{figure}

Before closing, we estimate the difference in the calculated number of neutrino-floor events between the conventional Helm-type and DSM predictions by defining the ratio 
\begin{equation}
\mathcal{R}=\frac{\mathrm{DSM_{events}}}{\mathrm{Helm_{events}}} \, .
\end{equation}
For each nuclear system the corresponding results are presented in Fig.~\ref{fig:DSM_over_Helm} indicating that the differences can become significant, especially in the high energy tail of the detected recoil spectrum.
\section{Conclusions}
\label{sect:conclusions}

In this work, we studied comprehensively the expected event rates in WIMP-nucleus and neutrino-floor processes by performing reliable calculations for a set of prominent nuclear materials of direct dark matter detection experiments. The detailed calculations involve crucial nuclear physics inputs in the framework of the deformed shell model based on Hartree-Fock nuclear states. This way, the nuclear deformation as well as the spin structure effects of odd-$A$ isotopes, that play significant role in searching for dark matter candidates, are incorporated. The chosen nuclear detectors involve popular nuclear isotopes in dark matter investigations such as the $^{71}$Ga, $^{73}$Ge, $^{75}$As and $^{127}$I isotopes. The DSM results indicate that $^{71}$Ga needs further investigation by employing another effective two-body interaction than the one used in the chosen set of nuclear isotopes. 

The deformed shell model (DSM) employed for the nuclear structure calculations in this work is very well tested in many examples in the past~\cite{ks-book} for nuclei with $A$=60--90. Therefore, in our study we have chosen the dark matter candidates $^{71}$Ga, $^{73}$Ge and  $^{75}$As. In addition, to extend DSM to heavier nuclei of interest in dark matter detection, we have considered $^{127}$I and the results, reported in the present paper are quite encouraging. In the near future we will consider Xe isotopes that are also of current interest. For lighter candidate nuclei, such as Na, Si and Ar, clearly shell model will be better choice and DSM may also be tested for these isotopes.

More importantly, by exploiting the expected neutrino-floor due to Solar, Atmospheric and DSNB neutrinos, which constitute an important source of background to dark matter searches, the impact of new physics CE$\nu$NS contributions based on novel electromagnetic neutrino properties and $Z^\prime$ mediator bosons have, been estimated and discussed. Our results also indicate that the addressed novel contributions may lead to a distortion of the expected recoil spectrum that could limit the sensitivity of upcoming WIMP searches. Such aspects could also provide key information concerning existing anomalies in $B$-meson decays at the LHCb experiment~\cite{Dalchenko:2017shg}, and offer new insights to the LMA-Dark solution~\cite{Farzan:2015doa,Coloma:2017egw}. 

Finally, the present results indicate that the addressed nuclear effects may become significant, leading to alterations especially in the high energy tail of the expected neutrino-floor as described by effective nuclear calculations, thus motivating further studies in the context of advanced nuclear physics methods such as the deformed shell model or the Quasiparticle Random Phase approximations and others. Such a comprehensive study using available data of the COHERENT experiment is under way and will be presented elsewhere.

\section*{Data Availability}
The data used to support the findings of this study are available from the corresponding author upon request.
\section*{Conflicts of Interest}
The authors declare that they have no conflicts of interest.

\section*{Acknowledgements}
R. Sahu is thankful to SERB of Department of Science and Technology (Government of India) for financial support. DKP is grateful to Prof. Naumov for stimulating discussions.

\bibliographystyle{apsrev4-1}

%

\end{document}